\documentclass{aa}  

\usepackage{graphicx}
\usepackage{txfonts}
\usepackage{lipsum}
\usepackage{subcaption}         
\usepackage{lscape}             
\usepackage{placeins}           
\usepackage{multirow}

\usepackage{hyperref}
\hypersetup{
    colorlinks=true,
    citecolor=blue,
    linkcolor=blue,
    urlcolor=blue
}

\usepackage{float}

\begin{document}

    \title{Gen-EBM: an energy balance model for terrestrial planets}

   \subtitle{Climate multistability from snowball to post-runaway greenhouse climates}

   \author{Siddharth Bhatnagar\inst{1,2,3}\corrauth{siddharth.bhatnagar@unige.ch}        
        \and Emeline Bolmont\inst{1,3}\email{emeline.bolmont@unige.ch}
        \and Maura Brunetti\inst{2,3}\email{maura.brunetti@unige.ch}
        \and Jérôme Kasparian\inst{2,3}\email{jerome.kasparian@unige.ch}
        \and Tim Kallioinen\inst{1,3}\email{tim.kallioinen@unige.ch}
        \and Marine Leyvraz\inst{4,5}\email{marine.leyvraz@unibe.ch}
        \and Guillaume Chaverot\inst{6}\email{guillaume.chaverot@univ-grenoble-alpes.fr}
        \and Yonatan Birhanu Haile\inst{1}\email{yonatan.haile@etu.unige.ch}
        \and Dóra Takács\inst{1}\email{dora.takacs@etu.unige.ch}
        }

   \institute{Observatoire astronomique de l’Université de Genève, Versoix, Switzerland
   \and Group of Applied Physics and Institute for Environmental Sciences, University of Geneva, Geneva, Switzerland
   \and Centre pour la vie dans l'Univers de l'Université de Genève, Genève, Switzerland
   \and Climate and Environmental Physics, University of Bern, Bern 3012, Switzerland
   \and Oeschger Centre for Climate Change Research, University of Bern, Bern 3012, Switzerland
   \and Univ. Grenoble Alpes, CNRS, IPAG, F-38000 Grenoble, France}

   \date{Received September 30, 20XX}
 
  \abstract
  {}
   {
   Climate multistability and tipping points are fundamental properties of planetary climates that depend sensitively on the underlying climate feedbacks. We introduce the Geneva Energy Balance Model (Gen-EBM), a new 1-D latitudinal EBM designed for rapid exploration of terrestrial planet climates, and use it to investigate how planetary rotation, eccentricity, and thermal inertia shape climate attractors and multistability.} 
   {
   Gen-EBM is implemented in Rust and builds upon previous exoplanet EBMs by incorporating a temperature-dependent outgoing longwave radiation prescription spanning Earth-like, runaway-onset, and post-runaway greenhouse regimes. We construct bifurcation diagrams from ensembles of simulations and explore the effects of rotation rate, orbital eccentricity under the mean-flux approximation, and surface thermal inertia.}
   {
   The inclusion of post-runaway greenhouse states substantially enlarges the region of parameter space over which multiple climate states coexist. Within our model framework, slow rotators approach the spatially homogeneous limit and maintain broader temperate stability ranges, whereas fast rotators develop strong equator-to-pole temperature gradients that narrow the temperate branch. Under the mean-flux approximation, increasing eccentricity enhances climate variability and can modify the attractor structure itself. The critical eccentricity at which temperate climates disappear depends strongly on planetary thermal inertia, with ocean-rich planets remaining stable to larger eccentricities than land-dominated planets.}
   {
   Our results highlight the central role of climate feedbacks in shaping planetary climate attractors. Building on the long tradition of exoplanet EBMs, Gen-EBM broadens rapid parameter-space exploration to climates spanning temperate, runaway-onset, and post-runaway regimes. Its computational efficiency and flexibility make it a valuable bridge between simple climate models and 3-D general circulation models.}


   \maketitle
   \nolinenumbers

\section{\label{sec:introduction}Introduction}

The discovery of thousands of exoplanets, including terrestrial planets spanning a wide range of orbital distances, eccentricities, rotation rates, stellar hosts, and surface conditions \citep{leger2004new,miguel2010planet,winn2015occurrence}, has ushered in a new era in the study of planetary climates \citep[e.g.,][]{forget2014possible}.
A central objective of exoplanet science is to determine whether planets can sustain surface liquid water and are therefore potentially habitable \citep{kasting1993}. Since observations of terrestrial exoplanets remain limited, even with facilities such as the James Webb Space Telescope \citep[JWST,][]{gardner2023james}, numerical climate models play a vital role in interpreting present and future observations.

Habitability is not determined solely by stellar irradiation but also by climate feedbacks \citep[e.g.,][]{meadows2018,bhatnagar2026climates}.  
A striking example is the positive ice-albedo feedback \citep[e.g.,][]{pierrehumbert2010}, which is thought to have contributed to episodes of global or near-global glaciation on Earth \citep{evans1997,hoffman1998,hoffman2002,pierrehumbert2011}.
At the opposite extreme lie the moist and runaway greenhouse limits \citep{kasting1988,kasting1993}, which define the inner edge of the habitable zone \citep{kasting1993}. As surface temperatures increase, the concentration of atmospheric water vapour rises, strengthening the greenhouse effect until the atmosphere becomes increasingly opaque in the infrared (IR).\footnote{The runaway greenhouse onset is commonly linked to the Komabayashi-Ingersoll limit \citep{komabayasi1967,ingersoll1969} and the Simpson-Nakajima limit \citep{simpson1928,nakajima1992}.} As a result, the outgoing longwave radiation becomes unable to balance the absorbed stellar radiation, producing a positive radiative imbalance that drives rapid warming until a new post-runaway equilibrium is reached \citep[e.g.,][]{goldblatt2013low,leconte2013,goldblatt2015,turbet2021,chaverot2022does,chaverot2023first}. Depending on atmospheric composition and water inventory, these post-runaway states may attain temperatures of 1000--2000 K for an Earth-like ocean inventory \citep{kopparapu2013,turbet2021}. For Earth-like planets, this onset of runaway conditions is expected near 0.95 AU from the Sun \citep{kasting1993,leconte2013}. 
Besides these destabilising mechanisms, climate systems are also subject to stabilising feedbacks, such as the Planck feedback, namely the increase in outgoing longwave radiation with temperature \citep{pierrehumbert2010}.

The coexistence of competing feedbacks implies that a given external forcing need not correspond to a unique climate state, but may instead permit multiple stable solutions, or attractors \citep{milnor1985}. This climate multistability \citep{saltzman1983,ferreira2011, lucarini2014,brunetti2019} is a hallmark of nonlinear climate systems \citep{strogatz2018}. Transitions between attractors may occur through internal variability, bifurcations, or external shocks \citep{ashwin2012,brunetti2023}, potentially producing abrupt shifts between habitable and uninhabitable climates. Bifurcation diagrams provide a natural framework for studying such behaviour by identifying stable and unstable climate states, locating tipping points, and quantifying the extent of multistability.

Numerical climate models provide a powerful tool for exploring how climate multistability responds to planetary properties. At one end of the model complexity spectrum are three-dimensional (3-D) General Circulation Models (GCMs), which explicitly solve the governing equations of atmospheric and oceanic dynamics. Examples include the Generic-PCM \citep[][]{hourdin2006}, ROCKE-3D \citep{way2017resolving,tsigaridis2025rocke}, UM \citep{mayne2014,walters2017}, and ExoCAM \citep{wolf2022exocam}. While these models provide the highest degree of physical realism, their computational cost limits their suitability for broad parameter-space exploration.
This motivates more efficient approaches that retain key physical processes. One strategy is to develop physically grounded parameterisations in complex models, such as slab ocean models calibrated against comprehensive simulations \citep[e.g.,][]{bhatnagar2026fast}. Another is to reduce the spatial dimensionality of the climate system using simplified frameworks capable of rapidly surveying large regions of parameter space.

Energy balance models (EBMs) are one such class of reduced-complexity models. One-dimensional (1-D) latitudinal EBMs provide an efficient first-order description of planetary climate evolution \citep[e.g.,][]{north1979,spiegel2008}, and have long been used to study ice-line thresholds \citep{budyko1969,sellers1969}, climate sensitivity, long-term evolution, and multistability \citep[e.g.,][]{haqq2016,ramirez2020climate,simonetti2024seasonal}.

The efficiency of EBMs makes them particularly well suited to exoplanet applications, where large uncertainties in planetary properties necessitate broad parameter exploration. Exoplanet EBM studies have investigated the effects of meridional heat transport \citep{williams1997habitable}, rotation rate and ocean fraction \citep{spiegel2008}, obliquity and eccentricity \citep{spiegel2009,dressing2010,palubski2020habitability,wilhelm2022ice}, atmospheric composition and heat transport efficiency \citep{shields2013effect,vladilo2013habitable,vladilo2015modeling,haqq2016,ramirez2020effect}, and surface properties and their interaction with orbital forcing \citep{shields2013effect,ramirez2020climate,rushby2020effect,venkatesan2025one}. 
In the context of climate multistability, \citet{kadoya2014conditions} investigated the dependence of climatic states on insolation and CO$_2$ degassing, while \citet{checlair2017no} showed that synchronously rotating planets are unlikely to exhibit a snowball bifurcation. 
Collectively, these studies demonstrate the utility of EBMs for exploring planetary climate evolution.

However, most exoplanet EBMs have focused on climates spanning the snowball and temperate regimes and do not explicitly resolve the post-runaway greenhouse branch. 
Beyond EBMs, \citet{goldblatt2015} demonstrated using a radiative equilibrium model, that multiple stable states, including hot moist and very hot dry climates, can arise for pure-water atmospheres. 
\citet{palubski2020habitability} incorporated runaway greenhouse onset and water-loss limits in an EBM, but the evolution toward post-runaway equilibria was not explored. Capturing the full dynamical landscape of climate attractors and potential transitions between them therefore requires efficient models capable of spanning temperate, runaway-onset, and post-runaway regimes.

To address this need, we present the Geneva Energy Balance Model (Gen-EBM), a 1-D latitudinal EBM for rapid exploration of planetary climate parameter space. Building upon previous exoplanet EBMs, Gen-EBM incorporates an extended radiative formulation that enables the investigation of temperate \citep{spiegel2008}, runaway-onset \citep{chaverot2022does,chaverot2023first}, and post-runaway greenhouse climates \citep{turbet2021,chaverot2023first} within a unified framework.
Section~\ref{sec:model} describes the model and its physical parameterisations. Section~\ref{sec:results} investigates how planetary rotation, eccentricity and surface thermal inertia influence the stability of snowball, temperate, and post-runaway climate states. The implications and limitations of our results are discussed in Sect.~\ref{sec:discussion}, and Sect.~\ref{sec:conclusions} summarises our conclusions.

\section{\label{sec:model}Model}

\begin{figure*}[ht]
\sidecaption
\includegraphics[width=12.97cm]{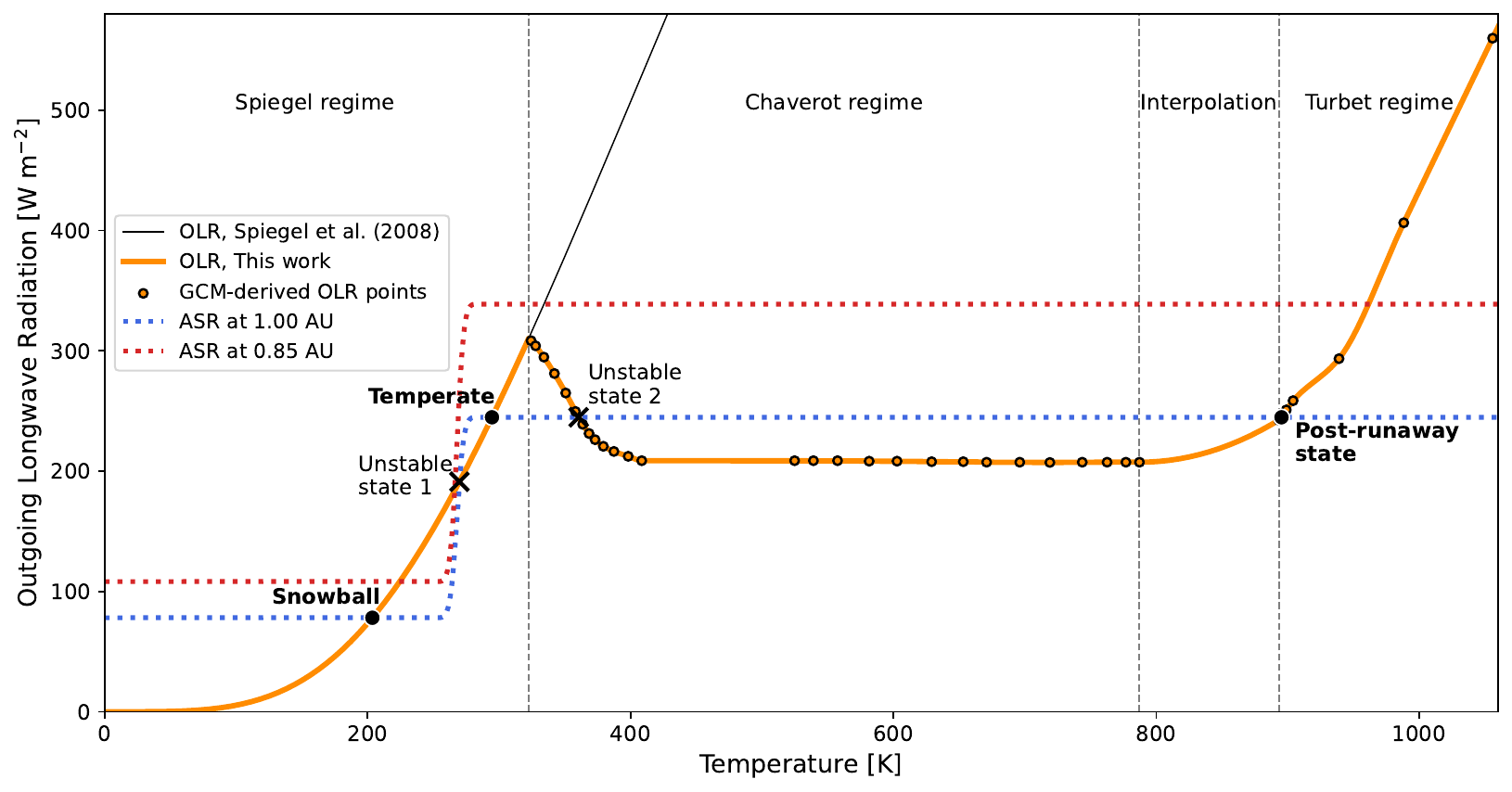}
\caption{Temperature-dependent OLR profile adopted in this work, based on \citet{spiegel2008}, \citet{turbet2021} and \citet{chaverot2023first}. At low temperatures, the profile follows the Earth-like OLR parameterisation of \citet{spiegel2008}. For $322.5~\mathrm{K} \lesssim T \lesssim 787~\mathrm{K}$, the OLR is based on \citet{chaverot2023first}, while for $893~\mathrm{K} \lesssim T < 1056~\mathrm{K}$, it follows \citet{turbet2021}. Dotted curves show the absorbed stellar radiation for a Sun-like star at orbital distances of 0.85 and 1.00 AU. For the 1.00 AU case, filled circles denote stable radiative equilibria and crosses unstable equilibria.}
\label{fig:predicting_attractors}
\end{figure*}

\subsection{\label{subsec:general-framework}General framework}

We use the Geneva Energy Balance Model (Gen-EBM), a 1-D latitudinal EBM for simulating the time-dependent climates of terrestrial planets. In this framework, the planet is represented by a zonally averaged meridional strip, where the prognostic variable is the surface temperature as a function of latitude and time. Gen-EBM builds upon the Williams-Kasting family of models \citep{williams1997habitable}, incorporating parameterisations from \citet{spiegel2008} and \citet{dressing2010}. Our implementation is written in Rust\footnote{Gen-EBM is publicly available at \url{https://github.com/DynaClim/ebm}.}, 
providing high computational efficiency (Appendix~\ref{app-subsec:computational-performance}) together with a modular framework that supports variable surface properties, radiative parameterisations, and orbital forcing configurations.

\subsection{\label{subsec:equation}Energy balance equation}
The model treats the meridional transport of heat as diffusion driven by the existing temperature gradient. The temperature $T$ at a given point $x$ (where $x=\sin{\lambda}$, and $\lambda$ is the latitude), evolves according to the diffusion-type energy balance equation of \citet{north1979}, \citet{north1981energy} and \citet{williams1997habitable}:
\begin{equation}
\label{eqn:EBM-equation}
C \frac{\partial T(x,t)}{\partial t}
- \frac{\partial}{\partial x} \left[ D(1 - x^2)\frac{\partial T(x,t)}{\partial x} \right]
+ I(T)
= S(x,t)\,\left(1 - A(T)\right),
\end{equation}
where $C$ is the effective heat capacity, $D$ is the diffusion coefficient describing the efficiency of meridional heat transport, $I(T)$ is the infrared cooling term (outgoing longwave radiation, OLR), $S$ is the diurnally averaged instellation and $A(T)$ is the planetary albedo. 
The net radiative balance is determined by the energy source term, the absorbed stellar radiation (the right-hand-side of Eq.~\ref{eqn:EBM-equation}), and the sink term, the OLR ($I(T)$). 
The governing equation is integrated numerically as described in Appendix~\ref{app-subsec:model-integration}.
The parameterisations adopted for these quantities are described in the following sections.

\subsection{\label{subsec:radiative-parameterisations}Radiative parameterisations}

\subsubsection{\label{subsubsec:OLR-function}Outgoing longwave radiation}

The IR-cooling term, $I(T)$, is prescribed using a composite temperature-dependent OLR parameterisation spanning snowball, temperate, runaway-onset, and post-runaway climates:
\begin{equation}
I(T)=
\begin{cases}
I_{\rm Spiegel}(T), & T<322.5~{\rm K},\\
I_{\rm GCM}(T), & 322.5~{\rm K}\le T<1056~{\rm K},
\end{cases}
\label{eqn:OLR-piecewise-equation}
\end{equation}
where the low-temperature branch follows "Model~2" of \citet{spiegel2008},
$I_{\rm Spiegel}(T)=\sigma T^4/\left[1+0.5925\left(\frac{T}{273\:{\rm K}}\right)^3\right]$,
where $\sigma$ is the Stefan--Boltzmann constant.
This parameterisation reproduces the greenhouse effect under present-day Earth-like conditions reasonably well \citep{spiegel2008}. 
The high-temperature branch, $I_{\rm GCM}(T)$, is obtained from GCM-derived OLR calculations spanning both the onset \citep{chaverot2023first} and post-onset phases \citep{turbet2021,chaverot2023first} of the runaway greenhouse.\footnote{The simulations used to derive these OLR values were performed with the Generic Planetary Climate Model (Generic-PCM; \citealt{hourdin2006}).} These calculations are particularly appropriate here because they were performed for Earth-like terrestrial planets, consistent with the atmospheric assumptions of our fiducial model.

This extended OLR formulation enables the model to resolve a rich palette of climate attractors, including snowball, temperate, and post-runaway states. Through Eq.~(\ref{eqn:OLR-piecewise-equation}), Gen-EBM is therefore well suited to investigate climate multistability and the broader range of climates expected among water-rich terrestrial exoplanets.

Since the OLR parameterisation encapsulates the radiative properties of the atmosphere, changes in atmospheric composition (e.g., CO$_2$ abundance or background gas inventory) would primarily be introduced through Eq.~(\ref{eqn:OLR-piecewise-equation}). Future versions of Gen-EBM will therefore incorporate composition-dependent OLR lookup tables derived from GCM simulations, enabling a more systematic exploration of atmospheric diversity.

Figure~\ref{fig:predicting_attractors} compares our adopted OLR profile (orange curve) with the original \citet{spiegel2008} parameterisation (black curve). 
Although the latter reproduces present-day Earth-like climates reasonably well, it cannot capture the onset of the runaway greenhouse, which is inherently a non-blackbody radiative process. 
To address this limitation, for $322.5~\mathrm{K} \lesssim T \lesssim 787~\mathrm{K}$, we adopt the OLR values of \citet{chaverot2023first}. 
They showed that the present-day Earth CO$_2$ abundance ($\approx$400 ppm) and continental distribution only weakly affect the climate evolution after the onset of the runaway greenhouse.
The initial surface water reservoir mainly controls the temperature at which the planet exits the runaway phase, corresponding to the subsequent rise in OLR.
The OLR parameterisation adopted for this temperature range is therefore applicable to CO$_2$ abundances below several thousand ppm and to water inventories ranging from $\sim$1 bar to several hundred bars.

For $893~\mathrm{K} \lesssim T < 1056~\mathrm{K}$, we use the post-runaway OLR values from \citet{turbet2021}. We interpolate between the two datasets to obtain a continuous OLR profile, shown by the orange curve in Fig.~\ref{fig:predicting_attractors}.
For post-runaway climates, both \citet{turbet2021} and \citet{chaverot2023first} showed that the OLR depends strongly on the initial surface water reservoir. The parameterisation adopted here (Fig.~\ref{fig:predicting_attractors}) is based on GCM simulations assuming an initial water inventory of 10 bars. Larger initial water inventories would shift the temperature at which the OLR rises after the runaway phase to higher values. Since GCM simulations of water-dominated atmospheres remain computationally expensive, extending the OLR lookup table to larger water inventories is left for future work.

The range of temperatures accessible to Gen-EBM is therefore ultimately constrained by the domain of the adopted OLR parameterisation rather than by the external forcing itself. In particular, temperatures exceeding $\approx 1056$ K lie beyond the upper limit of the OLR lookup table (Fig.~\ref{fig:predicting_attractors}). Since no extrapolation is performed beyond this limit, simulations that exceed this temperature are terminated and excluded from further interpretation.
At the other extreme, although temperatures below $\approx 200$ K can in principle be computed, caution is required when interpreting such states, as additional physical processes not included in Gen-EBM, such as atmospheric condensation or collapse, may become important \citep[e.g.,][]{venkatesan2025one}. Consequently, for the atmospheric composition assumed in the adopted OLR parameterisation, the practical range of applicability of Gen-EBM is approximately 200--1056 K.

\subsubsection{\label{subsubsec:absorbed-stellar-radiation}Absorbed stellar radiation}
The planetary albedo is prescribed using the temperature-dependent formulation of \citet{gilmore2014reliable}:
\begin{equation}
\label{eqn:Gilmore-albedo}
A[T] = \frac{a_l + a_h}{2} - \frac{a_l - a_h}{2}\tanh{\frac{T - T_c}{\Delta T}},
\end{equation}
where $a_l$ and $a_h$ are the low- and high-temperature asymptotic albedos, representing ice-covered and ice-free conditions, respectively, $T_c$ is the temperature at the centre of the transition, and $\Delta T$ determines its width. The hyperbolic tangent formulation provides a smooth transition between the ice-covered and ice-free albedo regimes.

Since surface albedo depends on the spectral distribution of incident radiation, Gen-EBM includes a two-channel (VIS/IR) treatment of ice, snow, and ocean albedos (Table~\ref{app-tab:albedo}). This is relevant for planets orbiting cool stars, for which the reduced near-infrared reflectivity of ice and snow weakens the ice-albedo feedback \citep[e.g.,][]{joshi2012suppression}.
The implications of this will be explored in a future investigation (see Sect.~\ref{subsec:future-directions}).

For Earth-like planets, we adopt the Model 2 parameters of \citet{spiegel2008}, which translate to the following values: $a_l \approx 0.77$, $a_h \approx 0.28$, $T_0=268$ K, and $\Delta T=5$ K. The resulting albedo function (orange line) is shown in Fig.~\ref{fig:heat_cap_albedo}.

The absorbed stellar radiation (ASR) is then computed from the diurnally averaged instellation, 
$S(x,t)$, following \citet{berger1978long} and \citet{hartmann1994}.
Gen-EBM permits arbitrary external forcing and supports non-zero eccentricity and obliquity, which may be specified as constants, prescribed time series, or sinusoidally varying functions.

\subsection{\label{subsec:heat-cap-diffusion}Heat capacity and meridional transport}
The effective heat capacity $C$ depends on the underlying surface type and serves as a proxy for thermal inertia. Following \citet{dressing2010}, different heat capacities are assigned to land, ocean, ice, and cold ice surfaces (Table~\ref{app-tab:heat-capacity}). 
Figure~\ref{fig:heat_cap_albedo} shows the dependence of both, the albedo (orange line) and the heat capacity (blue) with temperature.

Meridional heat transport is parameterised as diffusion, with the diffusion coefficient $D$ scaling inversely with the square of the planetary rotation rate \citep{williams1997habitable}:
\begin{equation}
\label{eqn:diffusion-equation}
D = D_{\text{fid}} \left(\frac{\Omega_p}{\Omega_{\oplus}}\right)^{-2},
\end{equation}
where $D_\text{fid}$ is the fiducial diffusion coefficient, taken here to be 0.4738 W\,m$^{-2}$\,K$^{-1}$, following the tuning procedure described in Appendix~\ref{app-sec:model-validation}, and $\Omega_p$ and $\Omega_{\oplus}$ are the rotation rates of the model planet and Earth, respectively.
This scaling reflects the reduced efficiency of meridional heat transport on rapidly rotating planets, where stronger Coriolis forces reduce latitudinal energy redistribution. Following \citet{spiegel2008} and \citet{dressing2010}, varying $D$ provides a simple framework for investigating the climatic effects of different rotation rates (Sects.~\ref{subsec:rotation-rate} and \ref{subsec:rotation-limitation}). The applicability of Eq.~(\ref{eqn:diffusion-equation}) may, however, be limited for very slowly rotating planets \citep{del1993equatorial,del1996simulations}.

\section{\label{sec:results}Results}

The model was validated against modern Earth climatology by tuning the fiducial diffusion coefficient to reproduce Earth's annual-mean temperature, zonal temperature profile and ice line position (see validation procedure in Appendix~\ref{app-sec:model-validation}).

\subsection{\label{subsec:baseline-climate-multistability}Baseline climate multistability}

\subsubsection{\label{subsubsec:radiative-equilibrium-states}Radiative equilibrium states}

Before integrating the model, the approximate location of climate attractors can be inferred from radiative balance arguments. Neglecting meridional heat transport, Eq.~(\ref{eqn:EBM-equation}) reduces to:
\begin{equation}
\label{eqn:predict_attractors}
    I(T)=S\left(1-A(T)\right),
\end{equation}
such that equilibrium states correspond to intersections between the OLR and ASR curves.

Figure~\ref{fig:predicting_attractors} illustrates this for two orbital distances. Varying the orbital distance vertically shifts the ASR profile, while the OLR remains unchanged because it depends only on the assumed atmospheric composition. Due to the sharp albedo transition near $263$--$273$ K, multiple intersections may occur for the same stellar forcing. For example, the 1.00 AU case exhibits five stationary states (highlighted in the figure).

However, solutions to a thermal balance equation such as Eq.~(\ref{eqn:predict_attractors}) are stable only if $\frac{d(Q^+ - Q^-)}{dT} < 0$, where $Q^+$ and $Q^-$ denote the heating (ASR) and cooling (OLR) rates, respectively \citep{spiegel2008}.  Essentially, stable solutions occur when small temperature increments cause cooling to exceed heating and when small decrements cause heating to exceed cooling.

The equilibrium structure obtained in Fig.~\ref{fig:predicting_attractors} arises from the interplay between the OLR (Sect.~\ref{subsubsec:OLR-function}) and ASR (Sect.~\ref{subsubsec:absorbed-stellar-radiation}) parameterisations. Importantly, the number and location of stationary states are model-dependent and may change if additional feedback mechanisms are included or if existing ones are modified. We refer the reader to Sects.~\ref{subsec:feedbacks} and \ref{subsec:comparison-with-literature} for more details.

\begin{figure*}[h]
\includegraphics[width=\hsize]{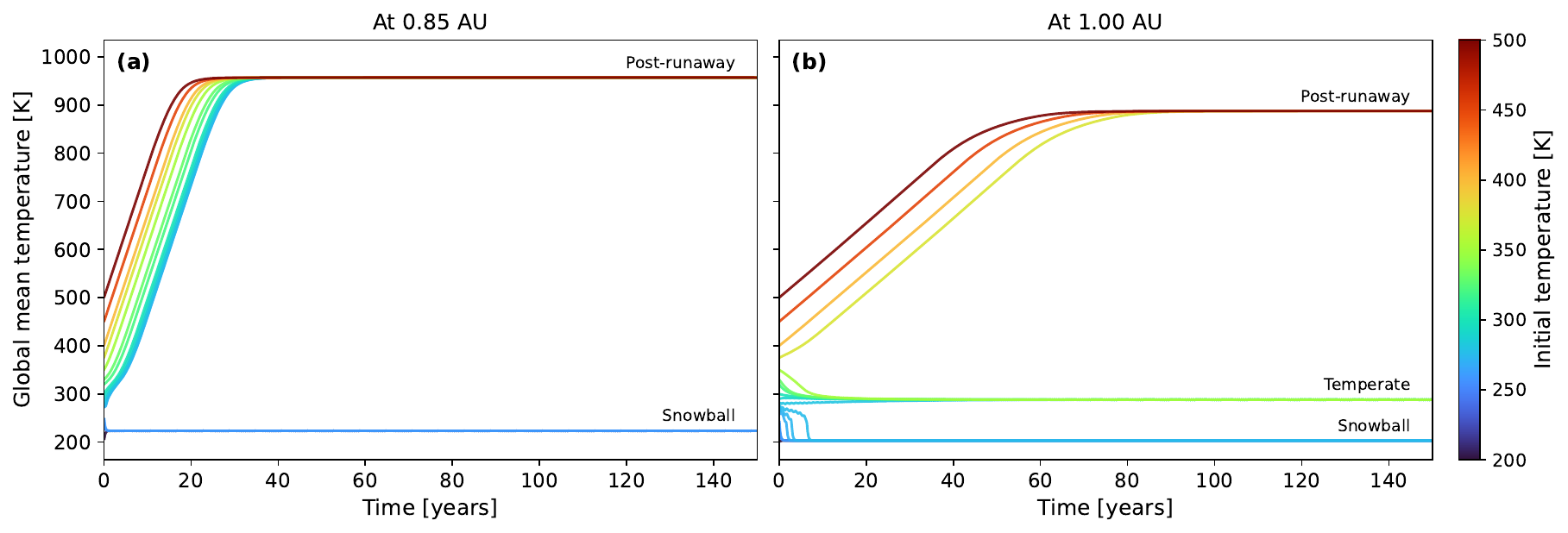}
\caption{Evolution of global mean temperature for the fiducial Earth-like model planet integrated from different initial temperatures: \textbf{(a)} At 1.00 AU from a Sun-like star, the model exhibits three stable states: Snowball ($203$ K), Temperate ($288$ K), and Post-Runaway ($888$ K). \textbf{(b)} At 0.85 AU, only two stable states remain: Snowball ($223$ K) and Post-Runaway ($950$ K).}
\label{fig:attractors-earth-obliquity}
\end{figure*}

\subsubsection{\label{subsubsec:the-attractors}Climate attractors}

To identify the stable states of the system, we integrate the fiducial Earth-like model from a range of isothermal initial conditions ($200\ \mathrm{K}\leq T_0\leq500\ \mathrm{K}$) and follow the evolution of the global mean temperature. Each simulation is integrated until equilibrium and the associated convergence times are detailed in Appendix~\ref{app-subsec:equilibration-time}.

Figure~\ref{fig:attractors-earth-obliquity} shows the thermal relaxation trajectories for two orbital distances. At 0.85 AU (panel a), the model exhibits two stable states: a snowball state ($223$ K) and a post-runaway state ($950$ K). The absence of a temperate attractor is consistent with the lack of a radiative equilibrium solution in the habitable temperature range for the 0.85 AU case (dotted red line in Fig.~\ref{fig:predicting_attractors}).

At 1.00 AU, we obtain three stable states (panel b): a snowball state ($203$ K), a temperate state ($288$ K), and a post-runaway state ($888$ K), in agreement with the radiative equilibrium analysis (dotted blue line in Fig.~\ref{fig:predicting_attractors}).
Given the lower stellar forcing at 1.00 AU, both the snowball and post-runaway states are cooler than their counterparts at 0.85 AU.

\begin{figure*}[h]
\centering
\includegraphics[width=\hsize]{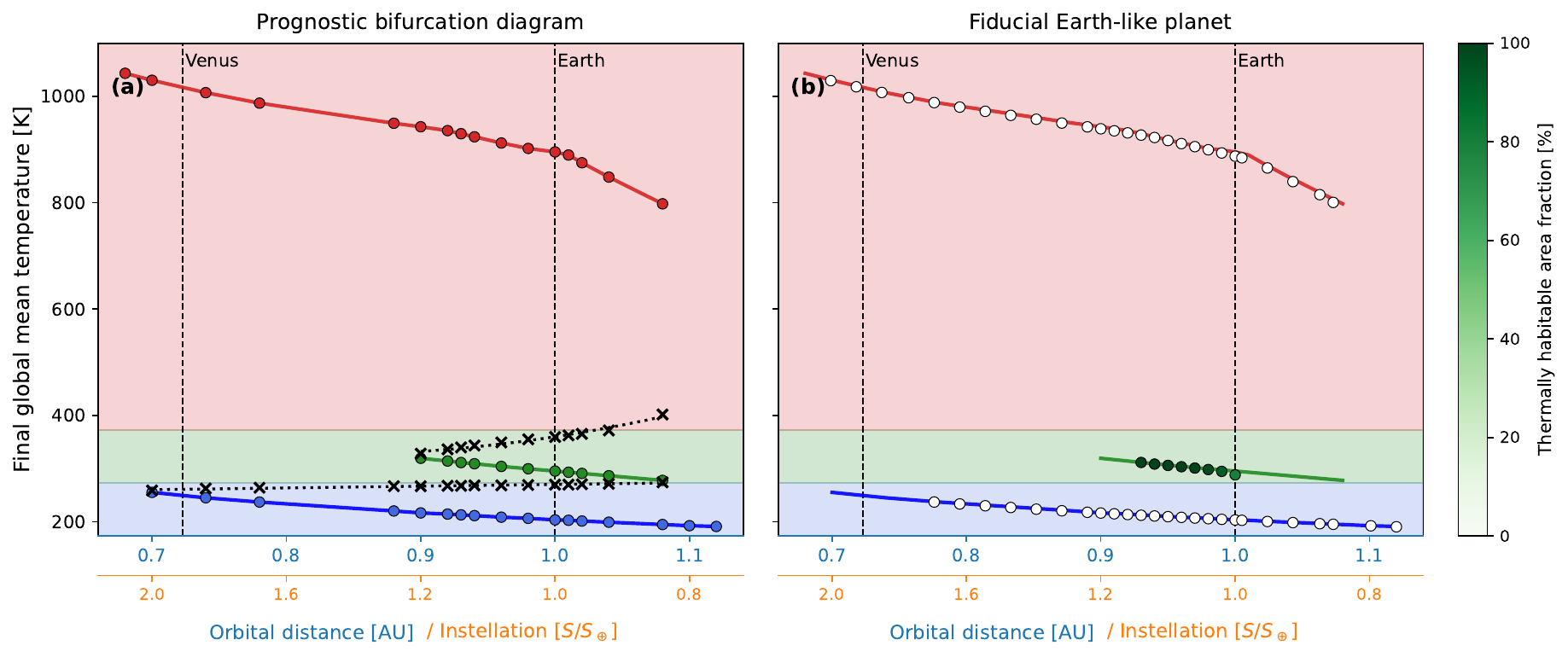}
\caption{Comparison between the prognostic and numerical bifurcation diagrams for the fiducial Earth-like configuration.
\textbf{(a)} Prognostic bifurcation diagram of the EBM. Red, green, and blue points denote post-runaway, temperate, and snowball states, respectively, while black crosses indicate unstable states. Shaded regions indicate thermally uninhabitable ($T>373$ K and $T<273$ K) and temperate ($273 \le T \le 373$ K) regimes.
\textbf{(b)} Numerical bifurcation diagram (decadally-averaged) of the fiducial Earth-like planet. The colour scale indicates the fraction of the planetary surface with temperatures between 273 and 373 K. White points denote thermally uninhabitable states. Stable branches from panel (a) are overplotted as solid curves.}
\label{fig:BD_prognostic_and_earth-obliquity}
\end{figure*}

\subsubsection{\label{subsubsec:bifurcation-structure}Bifurcation structure}

The thermal relaxation experiments can be extended over a range of stellar forcings to construct a bifurcation diagram, thereby identifying the stability ranges of the different climate states.
We first construct a "prognostic" bifurcation diagram by solving the local radiative equilibrium condition, Eq.~\ref{eqn:predict_attractors}, for each value of the external forcing.
Here, the equilibrium temperatures are obtained from the intersections between the ASR and OLR curves, without time-integrating the full latitudinal EBM.
Figure~\ref{fig:BD_prognostic_and_earth-obliquity}a shows the resulting prognostic bifurcation diagram, with stable and unstable branches indicated by solid and dotted lines, respectively.
This provides a first-order estimate of the equilibrium structure and allows unstable branches to be identified directly.

Figure~\ref{fig:BD_prognostic_and_earth-obliquity}b presents the numerical bifurcation diagram obtained  from 464 simulations arising from a combinatorial grid spanning 16 initial temperatures and 29 orbital distances between 0.68 and 1.12 AU. Mean temperatures are averaged over the final decade of each simulation.

The model predicts temperate climates between 0.93 and 1.00 AU, with an inner edge corresponding to an instellation of $\approx 390$ W m$^{-2}$, in close agreement with the $\approx 375$ W m$^{-2}$ threshold found by \citet{leconte2013} and \citet{chaverot2023first} using a 3-D GCM. The stable state branches from panel (a) have also been overlaid (solid lines) for reference. Compared to the prognostic bifurcation diagram, the extent of the temperate branch is reduced in the numerical model. The origin of these differences, which are linked to diffusion (rotation), is discussed in Sect.~\ref{subsec:rotation-rate}.

The location of the post-runaway branch depends on the adopted high-temperature OLR profile ($I_{\text{GCM}}(T)$ in Eq.~\ref{eqn:OLR-piecewise-equation}). In particular, larger initial water inventories would shift the rise of the OLR, and therefore the post-runaway branch, toward higher temperatures \citep{turbet2021,chaverot2023first}. By contrast, the temperate and snowball branches are expected to be comparatively less affected, since they are governed primarily by the low-temperature OLR ($I_{\text{Spiegel}}(T)$ in Eq.~\ref{eqn:OLR-piecewise-equation}) and ice-albedo feedback.

Snowball states exist between 0.78 and 1.07 AU, and post-runaway states extend outward to 1.07 AU. Consequently, there is a broad interval (0.78--0.93 AU) over which snowball states are possible even in the absence of temperate solutions, highlighting the strength of the ice-albedo feedback. More generally, this highlights how the temperate branch occupies a very narrow region in parameter space bounded by uninhabitable climates.

Figure~\ref{fig:BD_prognostic_and_earth-obliquity}b also reveals the (i) stability ranges of the different climate states, (ii) locations of bifurcations, and (iii) extent of climate multistability (0.78--1.07 AU), over which different attractors coexist and transitions between them are, in principle, possible. These properties are summarised in Table~\ref{tab:stability-ranges} and further discussed in Sect.~\ref{subsec:comparison-with-literature}.

The computational efficiency of Gen-EBM (Appendix~\ref{app-subsec:computational-performance}) enables a systematic exploration of how planetary properties can shape the bifurcation structure. In the following sections, we investigate three parameters particularly relevant to exoplanet diversity: planetary rotation rate (Sect.~\ref{subsec:rotation-rate}), and the effects of eccentricity and surface thermal inertia through variations in ocean fraction (Sect.~\ref{subsec:mean-flux-approximation}).

\begin{figure*}[h]
\centering
\includegraphics[width=\hsize]{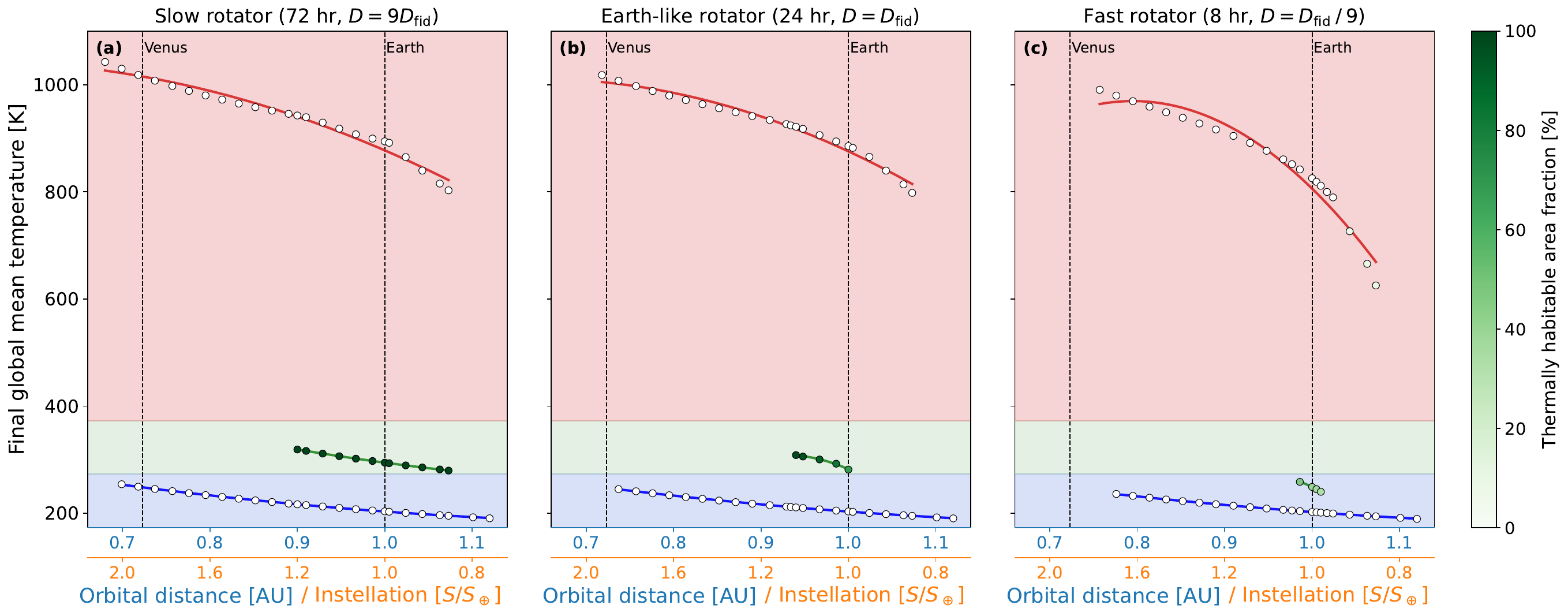}
\caption{Bifurcation diagrams for aquaplanets with rotation periods of \textbf{(a)} 72 hr, \textbf{(b)} 24 hr, and \textbf{(c)} 8 hr. Points denote decadal-averaged global mean temperatures of the climate attractors. Solid lines show second-order polynomial fits to the stable branches. Three post-runaway states in panel (c) are excluded from the fit (see Sect.~\ref{subsec:limitations}).}
\label{fig:BD-rotation-rates}
\end{figure*}

\subsection{\label{subsec:rotation-rate}Impact of rotation rate}
Planetary rotation is expected to vary substantially among terrestrial exoplanets, from rapidly rotating primordial states \citep[e.g.,][]{miguel2010planet} to slowly rotating / tidally evolved configurations \citep[e.g.,][]{goldreich1966spin}.
Slow rotators redistribute heat efficiently due to their weak Coriolis force. 
In the limit $\partial T/\partial x \rightarrow 0$, the climate approaches the spatially homogeneous assumption underlying the prognostic bifurcation diagram (Eq.~\ref{eqn:predict_attractors}). 
Conversely, rapid rotation strengthens the Coriolis force, weakening meridional transport, which promotes strong equator-to-pole temperature gradients. This would cause the climate attractors to deviate strongly from the prognostic prediction.

To investigate this effect, we repeat the bifurcation analysis for an Earth-like aquaplanet (100\% ocean coverage, zero eccentricity and obliquity), using the same combinatorial grid as in Sect.~\ref{subsubsec:bifurcation-structure}. We consider rotation periods of 72, 24, and 8 hr, corresponding via Eq.~(\ref{eqn:diffusion-equation}) to diffusion coefficients of $9D_{\rm fid}$, $D_{\rm fid}$, and $(1/9)D_{\rm fid}$, respectively.

Figure~\ref{fig:BD-rotation-rates} presents the resulting BDs. The slow-rotator case (72 hr, panel a) closely resembles the prognostic bifurcation diagram (Fig.~\ref{fig:BD_prognostic_and_earth-obliquity}a), confirming that efficient heat redistribution drives the climate toward the spatially homogeneous limit assumed in the prognostic treatment (Eq.~\ref{eqn:predict_attractors}; see also Fig.~\ref{fig:space-time-rotators-temperature}a). Relative to the Earth-like rotator (24 hr, panel b), the temperate branch extends over a substantially wider orbital range (0.90--1.07 AU vs. 0.95--1.00 AU).

\begin{figure*}[h]
\centering
\includegraphics[width=\hsize]{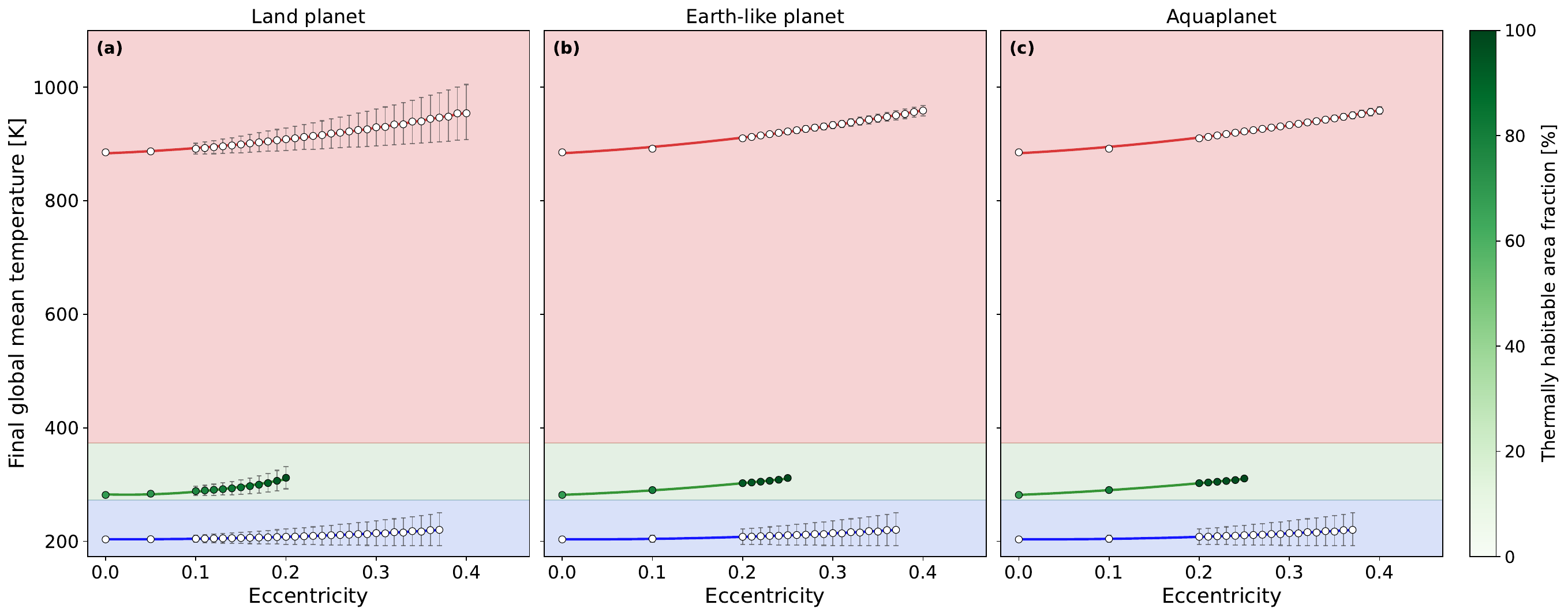}
\caption{Bifurcation diagrams as a function of eccentricity under the mean-flux approximation for \textbf{(a)} a land planet (90\% land, 10\% ocean), \textbf{(b)} an Earth-like planet (30\% land, 70\% ocean), and \textbf{(c)} an aquaplanet (100\% ocean). Points denote decadal-averaged climate attractors. Error bars indicate the minimum and maximum global mean temperatures during this period.}
\label{fig:BD-MFA-surfaces}
\end{figure*}

More generally, increasing the rotation rate systematically narrows the temperate regime (see Table~\ref{tab:stability-ranges}), as stronger equator-to-pole temperature gradients favour both snowball and post-runaway transitions. This behaviour is particularly pronounced for the fast rotator (8 hr, panel c), where weak meridional diffusion prevents excess equatorial heat from being efficiently redistributed to higher latitudes, making the planet more susceptible to transition into the post-runaway branch. Conversely, the inefficient transport of heat towards the poles allows high latitudes to cool more readily, which favours transitions to snowball conditions (see Sect.~\ref{subsec:feedbacks}). 
These strong temperature gradients also allow the temperate branch to retain non-zero liquid water area fractions (34--47\%), despite having globally averaged temperatures of only 240--259 K. This show how on fast rotators, warm equatorial regions can coexist with extensive icy mid-to-high latitudes (see Fig.~\ref{fig:space-time-rotators-temperature}c).

\subsection{\label{subsec:mean-flux-approximation}Impact of eccentricity and thermal inertia}

We adopt the mean-flux approximation \citep[MFA;][]{williams2002earth,bolmont2016habitability}, in which the semi-major axis is adjusted so that planets with different eccentricities receive the same orbit-averaged stellar flux as a reference circular orbit (here taken to be the Earth's annually averaged solar flux, $F_\oplus \approx 1360$ W\,m$^{-2}$). Under this assumption,
\begin{equation}
\label{eqn:mean-flux-approximation}
a = \frac{a_\oplus}{(1-e^2)^{1/4}}
\sqrt{\frac{L_s}{L_\oplus}},
\end{equation}
where the semi-major axis $a$ is written as a function of eccentricity $e$ and stellar luminosity $L_s$. Representative orbital parameters under the MFA are summarised in Table~\ref{app-tab:mfa-orbits}. Increasing eccentricity therefore leaves the orbit-averaged instellation unchanged while strengthening the periastron--apoastron contrast, allowing to isolate the effect of seasonal forcing on climate multistability.

For each eccentricity, corresponding to a unique MFA-adjusted semi-major axis (Eq.~(\ref{eqn:mean-flux-approximation})) and orbital period (Eq.~\ref{eqn:kepler-period}), we initialise Gen-EBM from the same ensemble of initial temperatures as in the previous bifurcation diagram experiments. Simulations are continued until either only a single uninhabitable attractor remains or the model reaches its upper temperature limit.

We consider three surface configurations with different effective heat capacities: a land planet (90\% land, 10\% ocean)\footnote{A small ocean fraction is retained to remain broadly consistent with the adopted OLR parameterisation (Fig.~\ref{fig:predicting_attractors}), which assumes an H$_2$O-saturated atmosphere. The limitations of this assumption are discussed in Sect.~\ref{subsec:limitations}. Nevertheless, Fig.~\ref{fig:BD_MFA_ocean_fraction_sensitivity} presents a sensitivity experiment comparing the nominal land planet (90\% land, 10\% ocean) with a fully land-covered planet.}, an Earth-like planet (30\% land, 70\% ocean), and an aquaplanet (100\% ocean). Since eccentricity primarily introduces a time-dependent forcing, the effective heat capacity controls the climatic response (following Table~\ref{app-tab:heat-capacity}).
The broad diversity of water inventories \citep[e.g.,][]{leger2004new} and eccentricities \citep[e.g.,][]{winn2015occurrence} expected among exoplanets further motivates our parameter choices.

Figure~\ref{fig:BD-MFA-surfaces} presents the resulting BDs for the three surface configurations. Error bars indicate the minimum and maximum global mean temperatures over the final decade of each simulation and therefore approximate the amplitude of orbital climate variability. As expected, temperature variability increases with eccentricity due to the growing periastron--apoastron flux contrast (Table~\ref{app-tab:mfa-orbits}), although its magnitude depends strongly on surface thermal inertia.

Let us first consider the temperate attractors. The land planet features a narrow temperate stability range, with no temperate state possible beyond $e \approx 0.20$ (panel a). Due to its low thermal inertia (Table~\ref{app-tab:heat-capacity}), the land planet responds rapidly to variations in stellar forcing, allowing temperatures near periastron to cross the runaway greenhouse threshold (Fig.~\ref{fig:predicting_attractors}).
In contrast, the Earth-like planet and aquaplanet (panels b and c) maintain stable temperate climates up to much larger eccentricities ($e \approx 0.25$) due to the strong thermal buffering provided by the ocean-dominated surface. As a result, the associated temperature variability is barely visible at the scale of the figure. 
This interpretation assumes that atmospheric radiative properties are independent of surface water inventory, a limitation discussed in Sect.~\ref{subsec:limitations}.

\begin{table*}[h]
\caption{\label{tab:stability-ranges}
Stability ranges of the climate attractors for the experiments considered in this work.}
\centering
\begin{tabular}{cccccc}
\hline\hline
\begin{tabular}[c]{@{}c@{}}Experiment\\type\end{tabular} &
\begin{tabular}[c]{@{}c@{}}Planet\\type\end{tabular} &
Post-runaway &
Temperate &
Snowball &
\begin{tabular}[c]{@{}c@{}}Climate\\multistability\end{tabular} \\
\hline

\multirow{1}{*}{\textbf{Fiducial$^{(a)}$}}
& Earth-like
& 0.70--1.07$^{(b)}$
& 0.93--1.00
& 0.78--1.12$^{(c)}$
& \textbf{0.78--1.07$^{(b,\:c)}$} \\

\hline

\multirow{3}{*}{\textbf{Rotation$^{(a)}$}}
& \begin{tabular}[c]{@{}c@{}}
72 hr aquaplanet
\end{tabular}
& 0.68--1.07$^{(b)}$
& 0.90--1.07
& 0.70--1.12$^{(c)}$
& \textbf{0.70--1.07$^{(b,\:c)}$} \\

& \begin{tabular}[c]{@{}c@{}}
24 hr aquaplanet
\end{tabular}
& 0.72--1.07$^{(b)}$
& 0.95--1.00
& 0.74--1.12$^{(c)}$
& \textbf{0.74--1.07$^{(b,\:c)}$} \\

& \begin{tabular}[c]{@{}c@{}}
8 hr aquaplanet
\end{tabular}
& 0.76--1.02$^{(b)}$
& 0.99--1.01
& 0.78--1.12$^{(c)}$
& \textbf{0.78--1.07$^{(b,\:c)}$} \\

\hline

\multirow{3}{*}{\textbf{Eccentricity$^{(d)}$}}
& Land
& 0.00--0.40$^{(e)}$
& 0.00--0.20
& 0.00--0.37
& \textbf{0.00--0.37} \\

& Earth-like
& 0.00--0.40$^{(e)}$
& 0.00--0.25
& 0.00--0.37
& \textbf{0.00--0.37} \\

& Aquaplanet
& 0.00--0.40$^{(e)}$
& 0.00--0.25
& 0.00--0.37
& \textbf{0.00--0.37} \\

\hline
\end{tabular}

\tablefoot{
\tablefoottext{a}{Orbital forcing values in AU.}
\tablefoottext{b}{The inner orbital distance limit is constrained by the OLR lookup table's maximum temperature ($\sim$1056 K).}
\tablefoottext{c}{Results below $\sim200$ K should be interpreted cautiously since the model does not feature atmospheric condensation or collapse.}
\tablefoottext{d}{Orbital forcing values in eccentricity.}
\tablefoottext{e}{Simulations were terminated once only a single attractor remained.}
}
\end{table*}

For the land planet, we find that the amplitude of the temperature oscillations is substantially larger than for the Earth-like planet and aquaplanet at the same eccentricity.
Also, for the snowball branch, states persist up to $e=0.37$, beyond which the post-runaway state is the only remaining attractor.
These features are directly linked to the lower thermal inertia of the land and ice surfaces (Table~\ref{app-tab:heat-capacity}), which allows the climate to respond more strongly to the enhanced periastron--apoastron forcing.

Interestingly, for a given eccentricity, the temperature variability of both, post-runaway and snowball states, is larger than that of temperate states (particularly apparent in panel a). This is again a consequence of thermal buffering. In the post-runaway regime, the surface is entirely in the low-heat-capacity land state (Fig.~\ref{fig:heat_cap_albedo}), while snowball climates are dominated by ice, whose effective heat capacity is lower than that of ocean-dominated temperate climates (see Fig.~\ref{fig:heat_cap_albedo} and Table~\ref{app-tab:heat-capacity}). Consequently, both branches exhibit larger orbital temperature variations than their temperate counterparts (see also Sect.~\ref{subsec:limitations}).

A notable result is that the Earth-like planet and aquaplanet BDs (panels b and c) are virtually indistinguishable. This suggests that, within our model framework, an Earth-like planet ($70\%$ ocean coverage) can be approximated reasonably well as an aquaplanet in terms of its global thermal response to eccentric forcing. The main difference lies in the equilibration timescale: the aquaplanet converges more slowly because of its larger effective heat capacity.

To assess the sensitivity of the results presented in Fig.~\ref{fig:BD-MFA-surfaces} to the adopted heat capacities, we repeated the MFA experiments by uniformly scaling all heat capacities by factors of 0.5 and 2 (Appendix~\ref{app-sec:thermal-inertia-sensitivity}). While the quantitative bifurcation thresholds and seasonal amplitudes depend on the assumed heat capacities, the broad trends discussed in this section remain unchanged.

\section{\label{sec:discussion}Discussion}

\subsection{\label{subsec:rotation-limitation}Planetary rotation as a control on climate multistability}
Rotation controls the efficiency of latitudinal heat transport and therefore the magnitude of equator-to-pole temperature gradients. 
Within our diffusive EBM framework, increasing the rotation rate systematically narrows the temperate branch (Fig.~\ref{fig:BD-rotation-rates}; Table~\ref{tab:stability-ranges}). Slow rotators approach the spatially homogeneous limit since efficient heat transport redistributes excess low-latitude energy toward colder regions, stabilising the climate against both runaway greenhouse transitions and global glaciation. Conversely, rapid rotators permit larger latitudinal temperature contrasts, reducing the range of stellar forcing over which globally temperate climates can persist.

\begin{figure*}[t]
\centering
\includegraphics[width=\hsize]{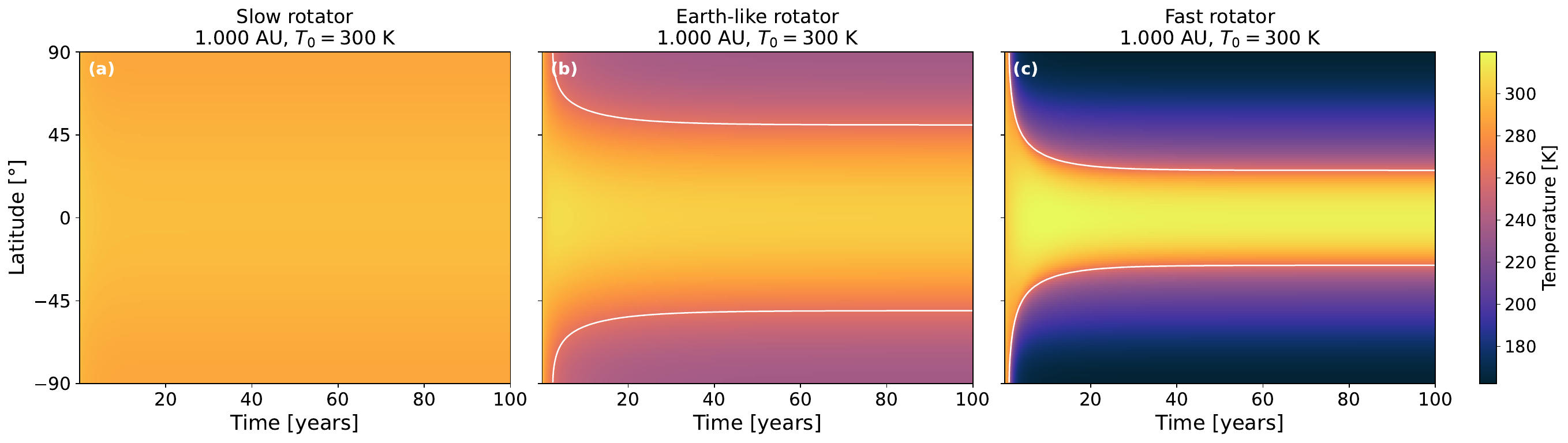}
\caption{Temperature space-time diagrams of the Earth-like aquaplanet at 1.000 AU with $T_0 = 300$ K for the three rotation-rate experiments. The white contour denotes the 263 K isotherm (see correspondence with the albedo space-time diagram, Fig.~\ref{fig:space-time-rotators-albedo}).}
\label{fig:space-time-rotators-temperature}
\end{figure*}

The fast-rotator experiments further highlight an important limitation of global mean temperature as a habitability metric. Despite globally sub-freezing mean temperatures, some fast-rotator temperate states retain substantial liquid-water area fractions (Fig.~\ref{fig:BD-rotation-rates}c), with warm equatorial regions coexisting with icy mid- and high-latitudes (Fig.~\ref{fig:space-time-rotators-temperature}c). 

The related consequences extend beyond temperature. The associated albedo patterns also differ substantially between rotation cases (Fig.~\ref{fig:space-time-rotators-albedo}), implying that planets receiving identical stellar fluxes may exhibit different observational signatures solely because of their rotation rate.

Figure~\ref{fig:space-time-rotators-temperature} illustrates the spatial origin of these differences. For identical forcing (1.000 AU and $T_0=300$ K), the slow rotator (panel a) remains nearly spatially homogeneous, the Earth-like rotator (panel b) develops a familiar ice-line structure (denoted by the 263 K isotherm), and the fast rotator (panel c) maintains a much narrower temperate belt with ice lines near $26.5^\circ$ latitude. Thus, diffusion (rotation) modifies not only the global mean temperature, but also its spatial distribution.

A useful comparison is provided by \citet{boschi2013bistability}, who used the intermediate-complexity GCM, PlaSim \citep{fraedrich2005plasim}, to study temperate--snowball bistability in solar-constant--CO$_2$ parameter space. They found that bistability persisted when the rotation rate was reduced by a factor of two, with only modest changes to the overall phase-space structure despite substantial circulation changes. Our results similarly show persistent multistability across rotation rates, but with a stronger sensitivity of the temperate branch width. This difference arises likely because rotation directly controls meridional diffusion in our EBM, whereas 3-D circulation can reorganise dynamically.

These results emphasise planetary rotation as a key control on not only habitability \citep{spiegel2009,dressing2010,meadows2018}, but also climate multistability. Two planets subjected to identical stellar forcing may occupy fundamentally different climate states solely because of differences in rotation rate. Moreover, assuming that atmospheric composition remains unchanged, secular spin-down \citep[e.g.,][]{goldreich1966spin} suggests that planetary bifurcation structures may evolve with time as rotation slows.

\subsection{\label{subsec:feedbacks}Feedbacks shape the attractors}
Change the feedbacks, and you change the attractors. The bifurcation structure obtained in this study arises from the competition between radiative and surface feedbacks. In particular, the ice-albedo and runaway greenhouse feedbacks determine the locations of the snowball, temperate, and post-runaway attractors.

The rapid transition of the warmest snowball trajectory in Fig.~\ref{fig:attractors-earth-obliquity} is illustrated more clearly in the albedo space-time diagrams of Fig.~\ref{fig:space-time-earth-obliquity-snowball}, shown for the 300 K initial condition but at two nearly identical orbital distances: 1.000 and 1.005 AU. 
Albedo serves as a proxy for ice cover through the temperature-dependent albedo prescription (Eq.~\ref{eqn:Gilmore-albedo}). At 1.005 AU, once the ice line reaches a critical latitude of $\approx 40^\circ$, the planet rapidly collapses into a snowball state -- a clear manifestation of the ice-albedo feedback. The negative slope of the transition through latitude-time space highlights the pole-driven nature of this instability. 
This existence of a critical ice-line threshold was first identified in pioneering 1-D EBM studies \citep{budyko1969,sellers1969}, further investigated by \citet{widiasih2013,walsh2013,deitrick2023functionality} and also in 3-D GCMs \citep[e.g.,][]{voigt2010transition,feulner2023tracing}.

The strength of the ice-albedo feedback also explains the remarkable resilience of snowball climates. Using an earlier version of our model, \citet{chaverot2024resilience} showed that even large perturbations, such as asteroid impacts and supervolcanic eruptions, are insufficient to destabilise a snowball state in the absence of long-term geologic feedbacks \citep{walker1981,kasting1993} and internal variability \citep{ashwin2012}. Similarly, we find that snowball climates remain stable as close as 0.78 AU in the fiducial Earth case (Fig.~\ref{fig:BD_prognostic_and_earth-obliquity}b) and even 0.70 AU for the slow rotator (Fig.~\ref{fig:BD-rotation-rates}a), corresponding to nearly twice the stellar flux received by modern Earth.

\begin{figure*}[h]
\centering
\includegraphics[width=\hsize]{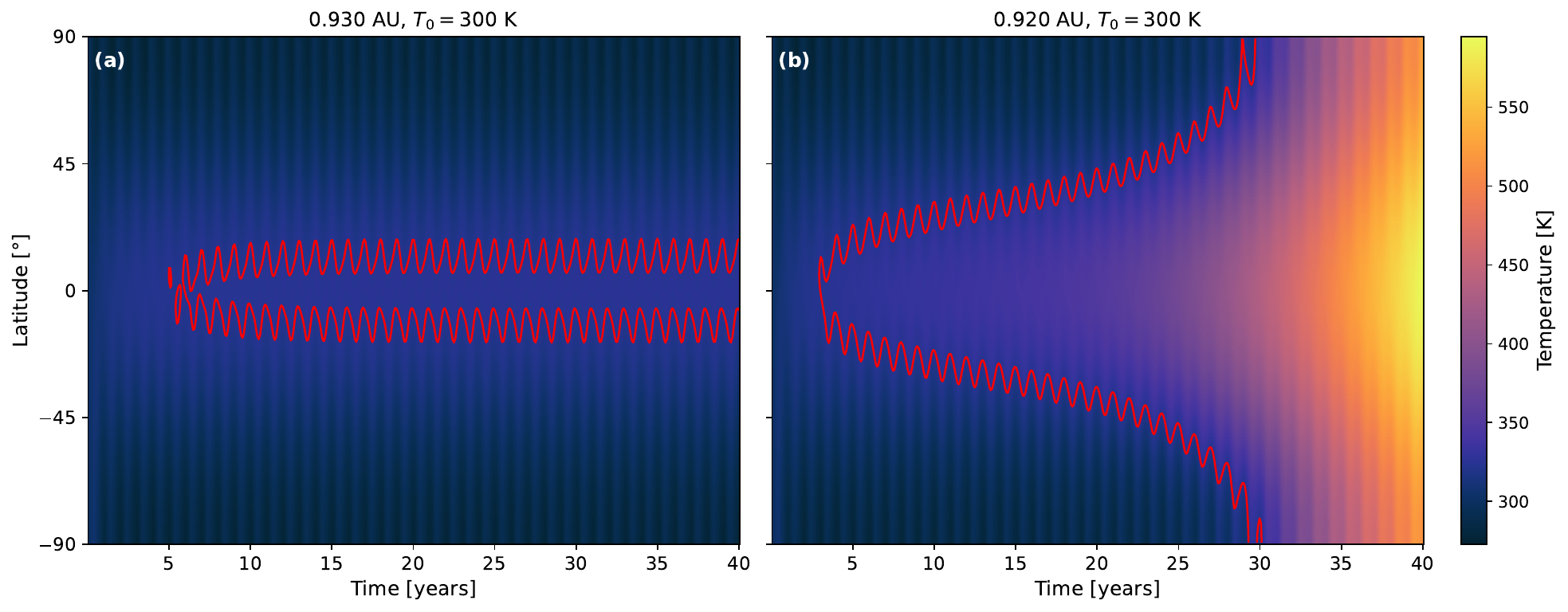}
\caption{Temperature space-time diagram for the 300 K initial condition for the fiducial Earth model planet at \textbf{(a)} 0.930 AU (temperate state), and \textbf{(b)} 0.920 AU (post-runaway). The red contour denotes the 323 K isotherm.}
\label{fig:space-time-earth-obliquity-runaway}
\end{figure*}

An important innovation of our EBM is the inclusion of a temperature-dependent OLR prescription that permits post-runaway states (Fig.~\ref{fig:predicting_attractors}). Figure~\ref{fig:space-time-earth-obliquity-runaway} illustrates the corresponding runaway transition in the form of a space-time diagram for the same initial condition as in Fig.~\ref{fig:space-time-earth-obliquity-snowball} ($T_0=300$ K). At 0.930 AU (panel a), the 323 K isotherm remains confined to low latitudes. At 0.920 AU (panel b), however, it propagates rapidly toward the poles, accompanied by a sharp increase in temperature.\footnote{The plots are truncated at 40 years so that the very high temperatures of the converged post-runaway state ($\approx 931$ K) do not dominate the colour scale.} Unlike the snowball transition, this instability is equator-driven, characterised by a positive slope in latitude-time space (also seen in 3-D GCM simulations by \citealt{chaverot2023first}).

Taken together, these results show that the dynamical landscape of climate attractors is governed primarily by the feedbacks included in the model. Although the precise locations of bifurcations may vary between models, the existence of competing stable states and abrupt transitions is a robust feature of planetary climates.

\subsection{\label{subsec:comparison-with-literature}Comparison to relevant EBM studies}
Firstly, due to our inclusion of a temperature-dependent OLR prescription permitting post-runaway greenhouse states (Fig.~\ref{fig:predicting_attractors}), for our fiducial Earth-like planet (Fig.~\ref{fig:BD_prognostic_and_earth-obliquity}b), the inner edge of the temperate branch lies at 0.93~AU ($\approx$ 390~W\,m$^{-2}$), in close agreement (within 4\%) with the $\approx 375$~W\,m$^{-2}$ threshold obtained with sophisticated 3-D GCMs by \citet{leconte2013} and \citet{chaverot2023first}. 
At 1.000 AU, our model reproduces a temperate state with a global mean temperature of $\approx 288$~K. However, the temperate branch disappears between 1.000 and 1.005 AU owing to the strong ice-albedo feedback. Similar behaviour was already found in other EBMs from the Williams-Kasting family: for instance, \citet{dressing2010} reported a transition to a snowball state when the stellar luminosity was reduced to $0.995~L_\odot$, while \citet{budyko1969} and \citet{hart1979} obtained comparable thresholds. However, in reality, this outer limit of the habitable zone is constrained by the carbonate-silicate cycle \citep[e.g.,][]{walker1981,kasting1993}.

More fundamentally, the inclusion of post-runaway greenhouse states in our EBM substantially modifies the dynamical landscape of the climate attractors (Figs.~\ref{fig:BD_prognostic_and_earth-obliquity}, \ref{fig:BD-rotation-rates} and \ref{fig:BD-MFA-surfaces}).
The existence of hot moist and very hot dry states was previously demonstrated by \citet{goldblatt2015} for pure-water atmospheres. Here, we show that incorporating post-runaway states within a latitudinal EBM framework similarly enlarges the range of climate multistability.
Traditional EBMs accounting only for the ice-albedo feedback predict snowball climates as the only stable state beyond the outer edge of the temperate branch. In contrast, our model permits post-runaway states out to $\approx 1.07$~AU, resulting in climate multistability over a much wider forcing range. For example, \citet{deitrick2023functionality} found multistability between $S/S_\oplus \approx 0.9$ and 1.35 using the VPlanet/POISE framework \citep{barnes2020vplanet}, whereas our fiducial Earth-like planet exhibits multistability over $S/S_\oplus \approx 0.87$--1.66 owing to the simultaneous inclusion of the ice-albedo and runaway greenhouse feedbacks (Table~\ref{tab:stability-ranges}). These results highlight two important aspects of climate multistability: first, it can occupy a remarkably large fraction of parameter space; and second, the number, location, and extent of climate states are fundamentally determined by the feedbacks included in the climate model.

Our eccentricity experiments also differ fundamentally from previous studies such as \citet{dressing2010} and \citet{spiegel2010generalized}, which considered planets at fixed semimajor axis. Under the mean-flux approximation \citep{williams2002earth,bolmont2016habitability}, increasing eccentricity modifies only the seasonal forcing while preserving the orbit-averaged stellar flux. Within this framework, and leveraging our OLR prescription (Fig.~\ref{fig:predicting_attractors}), we find that high eccentricities alter the underlying attractor structure, destabilising temperate climates through extreme periastron heating and favouring post-runaway states (Fig.~\ref{fig:BD-MFA-surfaces}). The critical eccentricity beyond which temperate states are no longer stable ($e \approx$ 0.10--0.25) depends strongly on the thermal inertia of the planet (Sect.~\ref{subsec:mean-flux-approximation}).

This behaviour contrasts with the results of \citet{bolmont2016habitability}, who found that highly eccentric planets around Sun-like stars tend to remain frozen because the short periastron passage is insufficient to melt the ice. A direct comparison is nevertheless difficult since those planets were in synchronous rotation. 
More generally, these differences illustrate that the climatic impact of eccentricity depends not only on the forcing itself, but also on the climate state, the feedbacks included in the model, and the spatial patterns that the model is able to resolve. In our case, these patterns are limited to one (latitudinal) dimension.

\subsection{\label{subsec:limitations}Model limitations}

Like all purely diffusive EBMs, Gen-EBM is fundamentally limited by its representation of heat transport through 1-D diffusion. While this captures the first-order effects of heat redistribution at low computational cost, it cannot explicitly represent large-scale atmospheric circulation and associated feedbacks. 
This limitation is illustrated by the fast-rotator experiments (Fig.~\ref{fig:BD-rotation-rates}), where three runaway-like attractors with global mean temperatures of 600--750 K still retain polar regions below the freezing point of water (Fig.~\ref{fig:space-time-fast-rotators-runaway}). These states arise because the reduced diffusion coefficient strongly limits poleward heat transport. 
In a realistic 3-D climate system, such a scenario would likely generate strong atmospheric circulation, latent heat transport, and cloud-radiative feedbacks \citep[see][]{chaverot2023first}, which would likely reduce these temperature contrasts. More sophisticated diffusion prescriptions that depend on relative humidity and heat capacity \citep[e.g.,][]{vladilo2015modeling} may also alleviate this behaviour. 
Nevertheless, the extent to which rapid rotation and the associated small Rossby deformation radius would preserve such gradients remains an interesting open question. Since no 3-D GCM study has explored the runaway greenhouse transition for rapidly rotating planets, this represents an interesting direction for future work.

Additionally, although our OLR prescription spans snowball, temperate, and post-runaway climates, it is based on GCM calculations for water-rich terrestrial planets \citep{turbet2021,chaverot2023first}. The amount and nature of background gas, continent distribution and water vapour inventory strongly influences both the runaway threshold and the temperature of the post-runaway equilibria \citep{chaverot2022does,chaverot2023first}. 
In particular, planets with depleted water reservoirs are expected to maintain drier atmospheres and therefore, OLR curves closer to blackbody-like behaviour (black curve in Fig.~\ref{fig:predicting_attractors}). This parameterisation is therefore not applicable to water-depleted planets, for which the runaway greenhouse behaviour may differ substantially. Indeed, \citet{abe2011} showed that for planets with severely depleted water reservoirs, tropical OLR values are above the traditional runaway limit as the air is unsaturated.

Finally, the effective heat capacities adopted in Gen-EBM follow \citet{williams1997habitable}, which were calibrated for temperate climates. While our high-temperature albedo limit (Fig.~\ref{fig:heat_cap_albedo}) is consistent with post-runaway bond albedo estimates (Extended Data Fig. 3 of \citealt{turbet2021}), optically thick steam atmospheres at $T_{\rm s}\sim800$--1000 K are themselves likely to constitute important thermal reservoirs that could increase the effective heat capacity of the climate. Therefore, while the existence of post-runaway attractors is robust within our framework, their variability amplitudes and equilibration timescales (Sect.~\ref{subsec:mean-flux-approximation}) should be interpreted with caution.

\subsection{\label{subsec:future-directions}Future directions}

Several extensions follow naturally from this work. For instance, while all bifurcation diagrams presented here vary a single control parameter, Gen-EBM could be used to construct multivariate diagrams, for example in orbital-distance--CO$_2$ parameter space, as done by \citet{boschi2013bistability} for the warm--snowball bistability. The efficiency of Gen-EBM (Appendix~\ref{app-subsec:computational-performance}) would allow such analyses to be performed rapidly over wide parameter ranges, enabling exploration of how the post-runaway greenhouse branch connects to temperate and snowball states.

Future versions should incorporate composition- and water-inventory-dependent OLR lookup tables from GCM studies (including but not limited to \citealt{turbet2021}, \citealt{chaverot2022does} and \citealt{chaverot2023first}). This would allow the effects of atmospheric diversity, background gas inventory, and surface water abundance on runaway onset and post-runaway equilibria to be explored more systematically.

Coupling Gen-EBM to orbital and tidal evolution models such as Posidonius \citep{blanco2017studying,bolmont2020impact,revol2024drifts,bolmont2025survival},\footnote{Gen-EBM and Posidonius share the same Rust-based architecture, see the DynaClim Github for more details: \url{https://github.com/DynaClim/}} would allow climate attractors to be followed along (long-term) evolving planetary trajectories, including tidal spin-down, eccentricity evolution, and Milankovitch-like forcing \citep[e.g.,][]{bolmont2016habitability,deitrick2018exo}.

Future implementations will also focus on obliquity, which was fixed in our rotation and mean-flux approximation experiments. Previous studies showed that it can strongly influence climate and ice distribution \citep{palubski2020habitability,wilhelm2022ice}.
Additionally, the physical parameterisations of Gen-EBM could be extended to include improved heat-transport scalings \citep[e.g.,][]{vladilo2015modeling,biasiotti2022eos}, more physically consistent albedo treatments separating surface and atmospheric contributions \citep{biasiotti2022eos}, and additional long-term feedbacks such as CO$_2$ condensation \citep{venkatesan2025one} and carbonate--silicate cycling \citep{walker1981,kasting1993,haqq2016}. These developments would broaden the range of planetary regimes accessible to Gen-EBM while preserving its role as a fast exploratory bridge between simple EBMs and 3-D GCMs.

Finally, Gen-EBM could also be adapted to synchronously rotating planets using tidally locked coordinates \citep{checlair2017no,haqq2022energy} or higher-dimensional EBMs \citep{ramirez2024new}. Participation in the FILLET protocol \citep{deitrick2023functionality,barnes2025functionality} within the CUISINES model intercomparison project \citep{sohl2024cuisines} will also enable direct comparison with established exoplanet EBMs.

\section{\label{sec:conclusions}Conclusions}

We investigated climate multistability with the Geneva Energy Balance Model (Gen-EBM), a Rust-based latitudinal EBM designed for rapid exploration of terrestrial climates. By extending previous exoplanet EBMs \citep{williams1997habitable} with an outgoing longwave radiation parameterisation spanning Earth-like \citep{spiegel2008}, runaway-onset \citep{chaverot2023first}, and post-runaway regimes \citep{turbet2021,chaverot2023first}, we explored climate attractors over a substantially broader temperature range than is typically considered. 
We obtain broad agreement with previous EBM and GCM studies for the onset of both runaway glaciation and runaway greenhouse regimes. The inclusion of post-runaway states considerably enlarges the range of climate multistability compared to traditional EBMs. In Gen-EBM, this broadened attractor structure results from the competition between the stabilising Planck feedback and the destabilising ice-albedo and runaway greenhouse feedbacks, highlighting how the feedbacks included in a climate model shape the dynamical landscape of climate attractors.

Within our diffusive EBM framework, planetary rotation exerts a strong control on climate stability. Slow rotators are nearly spatially homogeneous and maintain broad temperate stability ranges, whereas fast rotators develop strong equator-to-pole temperature gradients that substantially narrow the temperate branch. These differences may influence not only planetary habitability \citep[e.g.,][]{meadows2018}, but also exoplanet observability through their effect on planetary albedo.

Under the mean-flux approximation \citep[MFA;][]{williams2002earth}, the climatic response to eccentricity was found to depend strongly on planetary thermal inertia. Ocean-rich planets buffer seasonal forcing and maintain temperate climates to larger eccentricities than land-dominated planets. At sufficiently high eccentricities, extreme seasonal forcing can modify the attractor structure itself, destabilising snowball and temperate states and promoting transitions toward post-runaway climates. Our results further reinforce the idea that the validity of the MFA is not universal \citep{bolmont2016habitability} and depends on the climate state and planetary properties.

Future developments will focus on extending the atmospheric parameterisations, incorporating additional climate feedbacks, and coupling Gen-EBM to long-term orbital and rotational evolution models. 
As exoplanet observations continue to provide mostly limited constraints on planetary properties, broad exploration of climate parameter space \citep[e.g.,][]{paradise2022exoplasim,haqq2022sparse} remains essential. In this context, computationally efficient models such as Gen-EBM provide a valuable bridge between simple EBMs and 3-D GCMs by enabling systematic exploration of climate phase space and identifying regimes that warrant investigation with more physically comprehensive models.

\section*{Code availability}

The exact version of Gen-EBM used in this study, together with the configuration and setup files required to reproduce all experiments, is archived on Zenodo at \url{https://doi.org/10.5281/zenodo.22125342}. The development version of the source code is publicly available on GitHub at \url{https://github.com/DynaClim/ebm}.

\begin{acknowledgements}
      This work was carried out within the framework of the NCCR PlanetS, supported by the Swiss National Science Foundation (SNSF) under grants 51NF40\_182901 and 51NF40\_205606. 
      E.B. acknowledges the financial support of SNSF (grant number: 200021\_197176 and 200020\_215760). 
      M.B. acknowledges the financial support of SNSF (Sinergia Project No. CRSII5\_213539). 
      G.C. acknowledges the financial support of Centre National d'Études Spatiales (CNES).
      The authors are grateful to the referee and editor for their constructive comments.
      S.B. would like to thank the GAP-Nonlinearity and Climate team, the DynaClim team and Stéphane Goyette for helpful discussions. 
      The computations were performed at the University of Geneva's HPC clusters of Baobab and Yggdrasil. 
      This research has made use of NASA’s Astrophysics Data System. 
      We confirm that all figures included in this manuscript are original and created by the authors.
\end{acknowledgements}

\bibliographystyle{bibtex/aa}
\bibliography{paper}

\begin{appendix}
\nolinenumbers

\section{\label{app-sec:numerical-implementation}Numerical setup and performance}
\subsection{\label{app-subsec:model-integration}Model integration and evaluation}
The model is solved on a latitudinal grid of 146 points, where each grid cell represents the longitudinally averaged temperature at that latitude. Spatial derivatives are evaluated using finite differences, yielding a system of coupled ordinary differential equations based on Eq.~\ref{eqn:EBM-equation}. This system is then integrated in time using an explicit adaptive fifth-order Dormand--Prince Runge--Kutta scheme (DOPRI5).

At each timestep, Gen-EBM evaluates the absorbed stellar radiation (Sect.~\ref{subsubsec:absorbed-stellar-radiation}), outgoing longwave radiation (Sect.~\ref{subsubsec:OLR-function}), and diffusive transport terms (Sect.~\ref{subsec:heat-cap-diffusion}) before computing the temperature tendency at each latitude. An adaptive timestep algorithm with relative and absolute error tolerances of $10^{-14}$ ensures accurate resolution of both transient and equilibrium states. To facilitate analysis, outputs are written at regular intervals using high-order dense interpolation of the adaptive solution.

\subsection{\label{app-subsec:computational-performance}Computational performance}

The software architecture is designed to facilitate large parameter-space exploration. Configuration files are generated automatically from user-defined parameter ranges (including different orbital distances, initial temperatures, diffusion coefficients), allowing combinatorial sets of experiments to be constructed with minimal user intervention. Since each simulation is independent, the ensembles are embarrassingly parallel and can be distributed across an arbitrary number of processors, ideal for efficient construction of large bifurcation diagrams and multidimensional parameter studies.

At the standard spatiotemporal resolution adopted in this work, the total wall-clock time required to construct a bifurcation diagram (368--528 simulations) ranges from a few tens of seconds to several minutes, depending on the experiment. Table~\ref{tab:computational-performance} summarises the corresponding wall-clock times.

\begin{table}[ht]
\caption{\label{tab:computational-performance}
Wall-clock times for the different experiments.}
\centering
\begin{tabular}{cccc}
\hline\hline
\begin{tabular}[c]{@{}c@{}}Experiment\\type\end{tabular} &
\begin{tabular}[c]{@{}c@{}}Planet\\type\end{tabular} &
\begin{tabular}[c]{@{}c@{}}Number of\\simulations$^{(a)}$\end{tabular} &
\begin{tabular}[c]{@{}c@{}}Total\\runtime\\ $[$s$]^{(b)}$\end{tabular} \\ 
\hline

Fiducial
& Earth-like planet
& 464
& 557 \\

\hline

\multirow{3}{*}{Rotation}
&
\begin{tabular}[c]{@{}c@{}}
Slow-rotating\\
aquaplanet (72 hr)
\end{tabular}
& 464
& 974 \\

&
\begin{tabular}[c]{@{}c@{}}
Earth-like rotating\\
aquaplanet (24 hr)
\end{tabular}
& 464
& 158 \\

&
\begin{tabular}[c]{@{}c@{}}
Fast-rotating\\
aquaplanet (8 hr)
\end{tabular}
& 464
& 41 \\

\hline

\multirow{3}{*}{Eccentricity}
& Land planet
& 528
& 605 \\

& Earth-like planet
& 368
& 150 \\

& Aquaplanet
& 368
& 131 \\

\hline
\end{tabular}

\tablefoot{
$^{(a)}$ Number of simulations corresponds to the number of initial conditions and forcing combinations integrated for each experiment.\\
$^{(b)}$ Total wall-clock runtime measured using 24 CPU cores on a high-performance cluster (HPC).
}
\end{table}

We note that these runtime estimates are highly dependent on the computational resources available to the user. The values reported in Table~\ref{tab:computational-performance} were obtained using 24 CPU cores; shorter wall-clock times are possible when additional computational resources are available. For instance, we used 48 CPU cores for the MFA sensitivity analysis presented in Appendix~\ref{app-subsec:sensitivity-C-values}. The associated wall-clock times to construct the bifurcation diagrams for the land-, Earth-like and aquaplanet configurations then decreased to 259, 85 and 56 seconds respectively.

The model's computational efficiency stems from several factors. The code is implemented in Rust, arrays are preallocated and reused, and latitude-dependent trigonometric quantities are computed only once during initialization. Together with the 1-D nature of the diffusion problem, these optimisations enable rapid exploration of large climate parameter spaces while maintaining numerical accuracy.

A notable trend in Table~\ref{tab:computational-performance} is the increase in runtime with decreasing rotation period (i.e., increasing diffusion coefficient). 
The diffusion term is the stiffest part of Eq.~\ref{eqn:EBM-equation}.
Since stronger diffusion introduces shorter physical timescales into the governing equations, this forces the adaptive DOPRI5 integrator to adopt smaller timesteps to satisfy our prescribed error tolerance. Therefore, experiments with larger diffusion coefficients require more integration steps and longer runtimes.

\subsection{\label{app-subsec:equilibration-time}Equilibration times}
Previous EBM studies found that integrations of approximately 130 model years were generally sufficient to reach equilibrium for Earth-like climates \citep{spiegel2008,dressing2010}. However, equilibration times depend on both the thermal inertia of the system and the adopted radiative parameterisation. In particular, our extended OLR formulation (Sect.~\ref{subsubsec:OLR-function}) allows some runaway greenhouse trajectories to require several hundred years before reaching a steady state. We therefore integrate all simulations for at least 200 model years and extend individual runs where necessary to ensure convergence to equilibrium (in some cases up to 800 model years).

\section{\label{app-sec:add-model-details}Surface parameterisations}

\subsection{\label{app-subsec:spectral-albedo}Spectral dependence of surface albedo}

\begin{table}[ht]
\centering
\caption{\label{app-tab:albedo}Two-channel albedos (VIS/IR) for different surface compositions.}
\begin{tabular}{lccc}
\hline\hline
Wavelength range &
Ice\tablefootmark{a} &
Snow\tablefootmark{a} &
Open ocean\tablefootmark{b} \\
\hline
VIS & 0.682 & 0.978 & 0.07 \\
IR  & 0.134 & 0.418 & 0.06 \\
\hline
\end{tabular}
\tablefoot{
\tablefoottext{a}{Ice and snow albedo values from \citet{rushby2020effect}. VIS corresponds to $\lambda \in [0.2,0.78]~\mu$m and IR to $\lambda \in [0.78,12.2]~\mu$m.}\\
\tablefoottext{b}{Open-ocean albedo values from \citet{brandt2005surface}. VIS corresponds to $\lambda < 0.7~\mu$m and IR to $\lambda > 0.7~\mu$m.}
}
\end{table}

\clearpage

\subsection{\label{app-subsec:surface-heat-capacities}Surface heat capacities}

\begin{table}[h]
\caption{\label{app-tab:heat-capacity}Effective surface heat capacities used in the model (adapted from \citealt{williams1997habitable}).}
\centering
\begin{tabular}{lc}
\hline\hline
Surface type & Effective heat capacity \\
 & (J\,m$^{-2}$\,K$^{-1}$) \\
\hline
Land & $C_l = 5.24 \times 10^6$ \\
Ocean & $40\,C_l$ \\
Ice $(263\ \mathrm{K} < T < 273\ \mathrm{K})$ & $9.2\,C_l$ \\
Cold ice $(T < 263\ \mathrm{K})$ & $2.0\,C_l$ \\
\hline
\end{tabular}
\end{table}

\FloatBarrier

\subsection{\label{app-subsec:heat_cap_albedo}Heat capacity and albedo}

\begin{figure}[h]
\centering
\includegraphics[width=\linewidth]{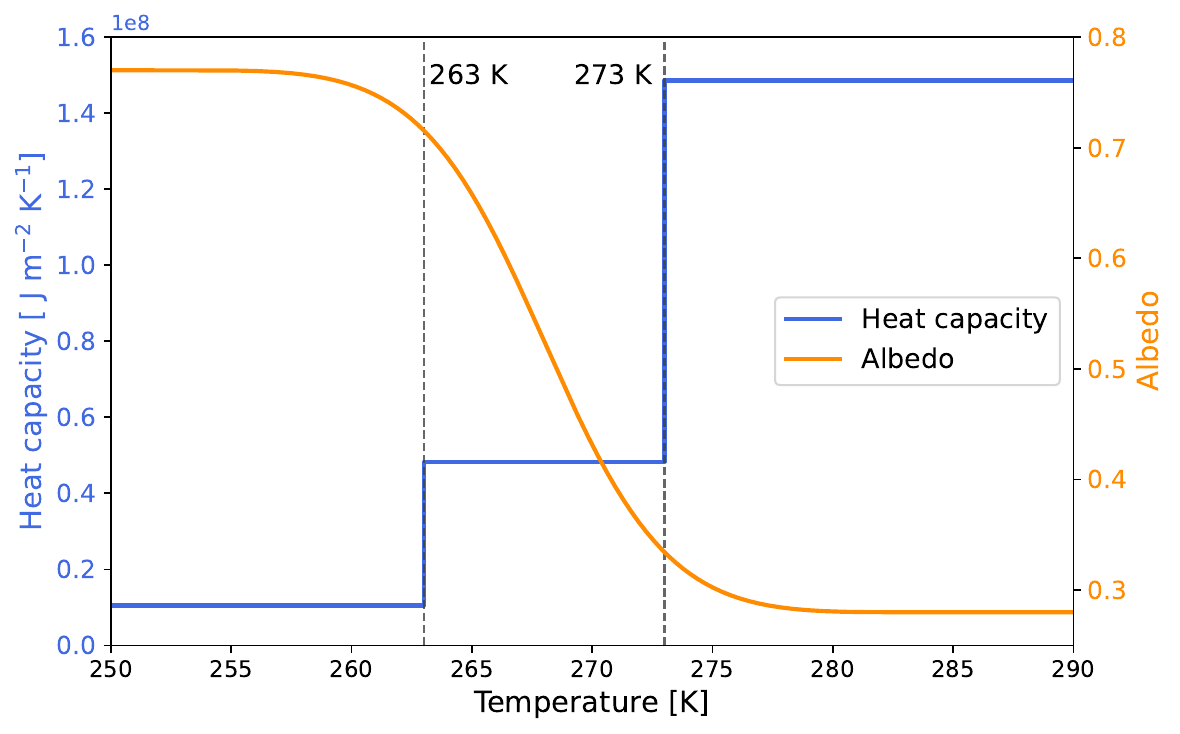}
\caption{Temperature-dependence of surface heat capacity and albedo in Gen-EBM. From a physical standpoint, these functions are proxies for the surface type on the planet.}
\label{fig:heat_cap_albedo}
\end{figure}

\FloatBarrier

\section{\label{app-sec:model-validation}Model validation}

\begin{figure}[ht!]
\centering
\includegraphics[width=\linewidth]{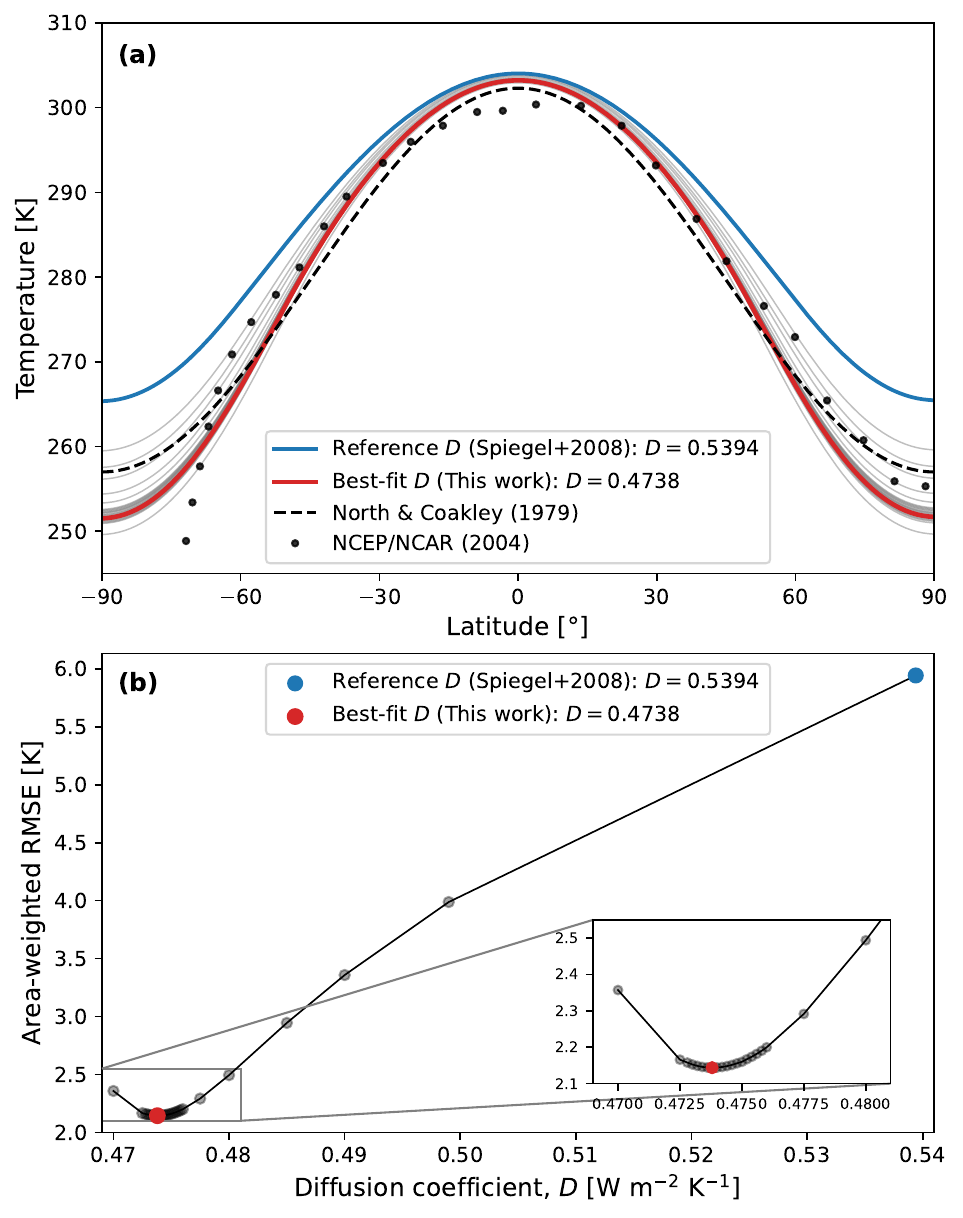}
\caption{Calibration of the diffusion coefficient for the fiducial Earth-like configuration.
\textbf{(a)} Annual-mean zonally averaged temperature profiles for Earth-like model configurations with different diffusion coefficients (grey lines), highlighting the reference diffusion coefficient (in W m$^{-2}$ K$^{-1}$) adopted by \citet{spiegel2008} (blue) and the best-fit value selected in this work (red). 
The dashed black line denotes the \citet{north1979} phenomenological fit, and black points show NCEP/NCAR reanalysis data \citep{kalnay1996}. \textbf{(b)} Area-weighted root-mean-square error (RMSE) relative to the \citet{north1979} profile as a function of diffusion coefficient $D$. The inset highlights the minimum RMSE region.}
\label{fig:model-validation}
\end{figure}

Modern Earth is our best-constrained example of a planetary climate and therefore serves as a natural benchmark for model validation. We define a fiducial Earth-like configuration with obliquity set to the present-day value (23.44$^\circ$), eccentricity to 0, and ocean fraction to 70\%. 
We then calibrate the diffusion coefficient, $D$, by comparing the resulting zonal temperature distribution to the phenomenological fit of \citet{north1979}, $T(\lambda) = 302.3\:K - 45.3\sin^2{\lambda}$, where $\lambda$ is the latitude. Figure~\ref{fig:model-validation} summarises this comparison.

Using the diffusion coefficient adopted by \citet{spiegel2008}, $D_{\mathrm{S08}}=0.5394$ W m$^{-2}$ K$^{-1}$, Gen-EBM produces temperatures that are systematically too warm, particularly at mid and high latitudes (Fig.~\ref{fig:model-validation}a, blue line). 
We therefore perform a sensitivity analysis over a range of diffusion coefficients (0.470--0.5394 W m$^{-2}$ K$^{-1}$; Fig.~\ref{fig:model-validation}b).\footnote{\citet{spiegel2008} also adopted a diffusion coefficient lower than that originally used by \citet{williams1997habitable}. Discrepancies between models likely arise from differences in numerical integration.}

We intentionally restrict the calibration to $D$ rather than performing a multi-parameter fit with surface albedo (Eq.~\ref{eqn:Gilmore-albedo}) or the IR tuning parameter (Eq.~\ref{eqn:OLR-piecewise-equation}). This is because our aim is to obtain a broadly Earth-like climate rather than reproduce every feature of the modern zonal temperature distribution. The \citet{north1979} profile (dashed black line in Fig.~\ref{fig:model-validation}a) provides a convenient phenomenological benchmark, while the NCEP/NCAR reanalysis \citep[][black points]{kalnay1996} acts as an observational reference. We therefore select the value of $D$ that minimises the area-weighted root-mean-square error (RMSE) relative to the \citet{north1979} profile while remaining broadly consistent with the reanalysis data and Earth-like ice-line latitudes. The area weighting also places greater emphasis on low- and mid-latitudes than on the polar regions.

The optimal agreement is obtained for $D=D_{\mathrm{fid}}=0.4738$ W m$^{-2}$ K$^{-1}$, which yields the minimum RMSE (Fig.~\ref{fig:model-validation}b, red point). This configuration produces a decadal-mean global surface temperature of $\approx 288$ K, in close agreement with the NCEP/NCAR value of $\approx 287$ K for the 1981--2010 period \citep{kalnay1996}, and places the seasonal-mean 263 K ice line near $65^\circ$ latitude. We therefore adopt $D=D_{\mathrm{fid}}$ throughout this study.

\clearpage

\section{\label{app-subsec:limitations-of-a-1-D-model}The limitations of a 1-D model}

\begin{figure}[ht!]
\centering
\includegraphics[width=\linewidth]{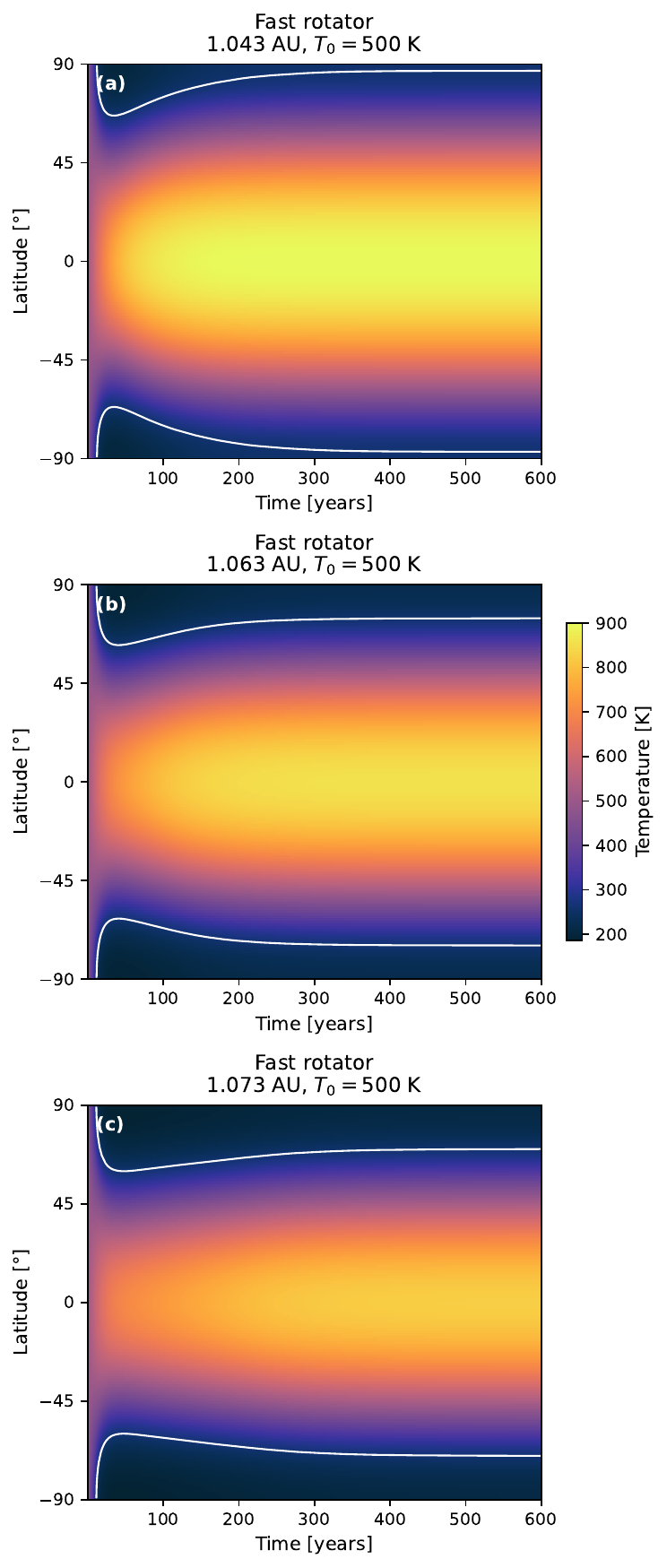}
\caption{Temperature space-time diagrams for the fast rotator at 1.043, 1.063 and 1.073 AU with $T_0 = 500$ K. The white contour denotes the 263 K isotherm.}
\label{fig:space-time-fast-rotators-runaway}
\end{figure}

\FloatBarrier

\section{\label{app-subsec:rotation-albedo}The impact of rotation on albedo}

\begin{figure}[ht!]
\centering
\includegraphics[width=\linewidth]{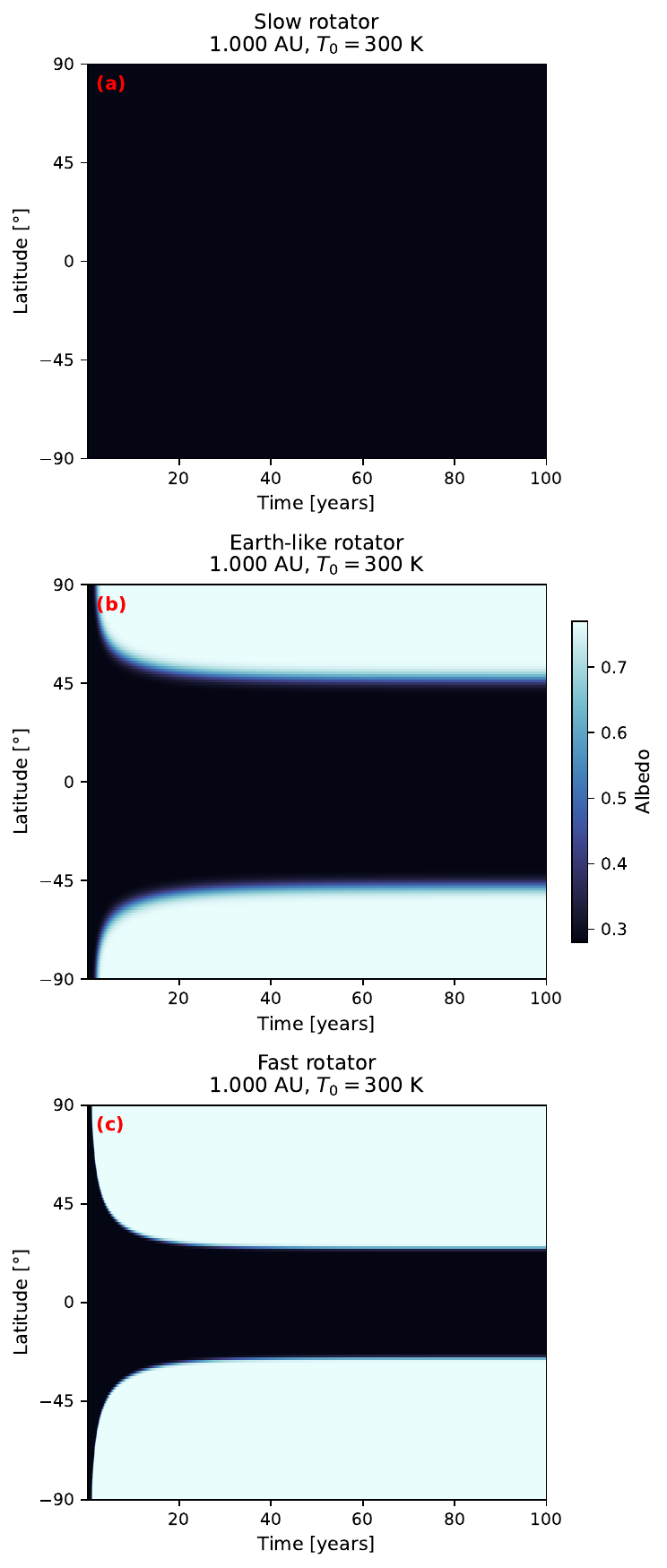}
\caption{Albedo space-time diagrams of the Earth-like aquaplanet at 1.000 AU with $T_0 = 300$ K for the three rotation-rate experiments (see correspondence with Fig.~\ref{fig:space-time-rotators-temperature}).}
\label{fig:space-time-rotators-albedo}
\end{figure}

\FloatBarrier

\section{\label{app-sec:snowball-transition}Space-time evolution of the snowball transition}

\begin{figure}[ht!]
\centering
\includegraphics[width=\linewidth]{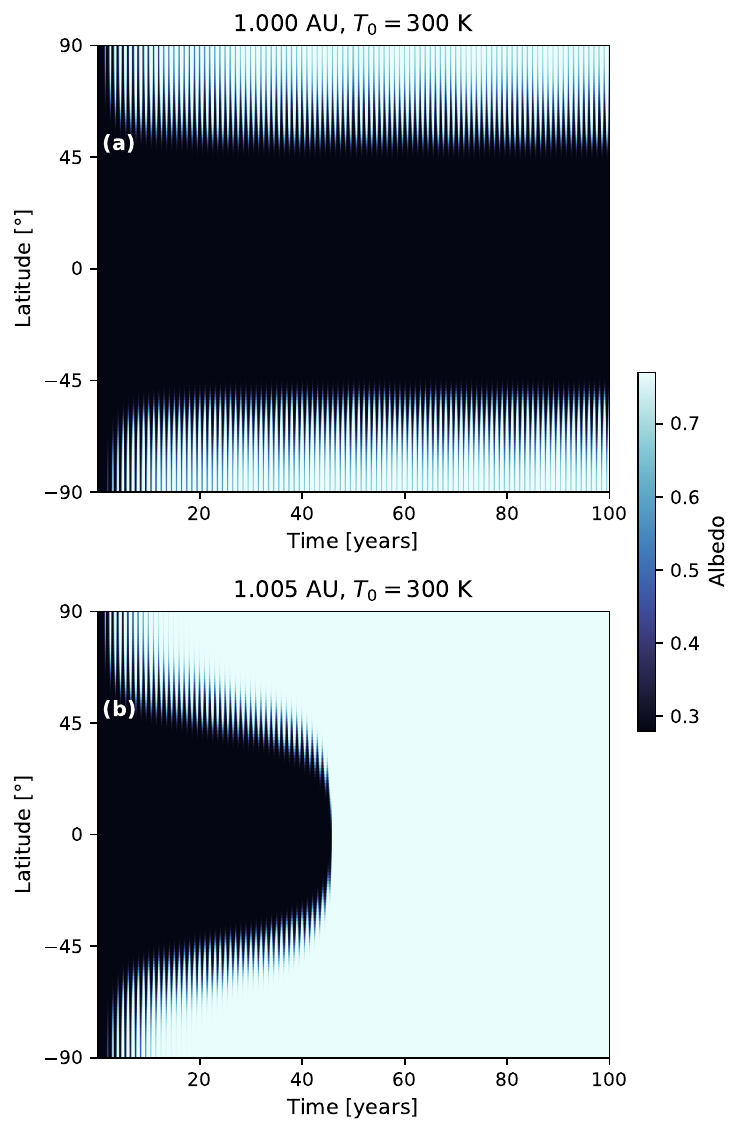}
\caption{Albedo space-time diagram for the 300 K initial condition for the fiducial Earth model planet at \textbf{(a)} 1.000 AU (temperate state), and \textbf{(b)} 1.005 AU (snowball).}
\label{fig:space-time-earth-obliquity-snowball}
\end{figure}

\FloatBarrier

\section{\label{app-sec:thermal-inertia-sensitivity}Mean-flux approximation parameters and sensitivity to thermal inertia}

\subsection{\label{app-subsec:orbital-parameters-MFA}Orbital parameters in the mean-flux approximation}

For the mean-flux approximation (MFA; Eq.~\ref{eqn:mean-flux-approximation}), the orbital period follows from Kepler's third law:
\begin{equation}
\label{eqn:kepler-period}
P_{\rm orb} = P_\oplus
\left(\frac{a}{a_\oplus}\right)^{3/2}
\left(\frac{M_\odot}{M_s}\right)^{1/2},
\end{equation}
where $M_s$ is the stellar mass. For the solar-type host star considered here, this reduces to $P_{\rm orb} = P_\oplus \, a^{3/2}$, where $P_\oplus=365$ days and $a$ is expressed in AU.
Table~\ref{app-tab:mfa-orbits} summarises selected eccentricity cases, together with the corresponding semi-major axis (Eq.~(\ref{eqn:mean-flux-approximation})), orbital period (Eq.~(\ref{eqn:kepler-period})), and stellar fluxes received at periastron and apoastron.

\begin{table}[h]
\caption{Orbital parameters and instantaneous stellar fluxes for representative eccentricities.}
\label{app-tab:mfa-orbits}
\centering
\begin{tabular}{cccccc}
\hline\hline
$e$ & $a$ & $r_{\rm peri}$ & $r_{\rm apo}$ & $P_{\rm orb}$ & $F_{\rm peri}$ / $F_{\rm apo}$ \\
 & (AU) & (AU) & (AU) & (day) & (W m$^{-2}$) \\
\hline
0.00 & 1.000 & 1.000 & 1.000 & 365.0 & 1360 / 1360 \\
0.11 & 1.003 & 0.893 & 1.113 & 366.7 & 1707 / 1097 \\
0.25 & 1.016 & 0.762 & 1.270 & 373.9 & 2341 / 843 \\
0.37 & 1.037 & 0.654 & 1.421 & 385.7 & 3183 / 673 \\
0.40 & 1.045 & 0.627 & 1.462 & 389.7 & 3462 / 636 \\
0.50 & 1.075 & 0.537 & 1.612 & 406.6 & 4711 / 523 \\
\hline
\end{tabular}
\end{table}

\subsection{\label{app-subsec:sensitivity-land-fraction}Sensitivity to land fraction}

\begin{figure*}[h]
\centering
\includegraphics[width=\hsize]{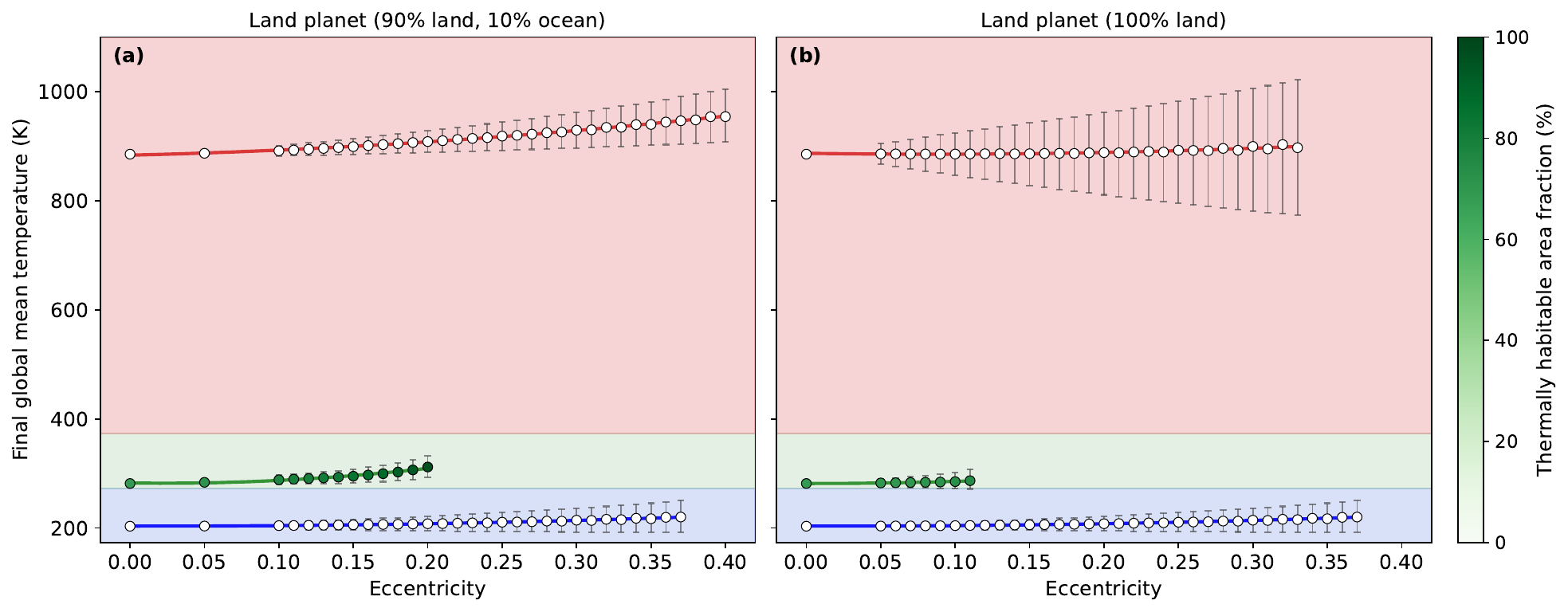}
\caption{Sensitivity to thermal inertia: bifurcation diagrams under the mean-flux approximation for \textbf{(a)} a land planet (90\% land, 10\% ocean) and \textbf{(b)} a fully land-covered planet (100\% land). Points denote decadal-averaged global mean temperatures of the climate attractors. Error bars indicate the minimum and maximum global mean temperatures during this period.}
\label{fig:BD_MFA_ocean_fraction_sensitivity}
\end{figure*}

Since thermal inertia emerges as an important factor controlling the validity of the MFA (see Sects.~\ref{subsec:mean-flux-approximation} and \ref{subsec:comparison-with-literature}), we perform an additional sensitivity experiment to isolate its influence.
For this, we compare the nominal land configuration (90\% land, 10\% ocean;~Fig.~\ref{fig:BD-MFA-surfaces}a) with an otherwise identical fully land-covered planet (100\% land). The corresponding bifurcation diagrams are shown in Fig.~\ref{fig:BD_MFA_ocean_fraction_sensitivity}. 
Although a completely land-covered planet is not expected to undergo a true water-vapour-driven runaway greenhouse, it still provides a useful limiting case for isolating the effect of substantially reduced thermal inertia.

Since all model parameters are identical apart from the ocean fraction, any differences between panels (a) and (b) can be attributed solely to the effective heat capacity of the climate. 
Figure~\ref{fig:BD_MFA_ocean_fraction_sensitivity} shows that even this modest reduction in thermal inertia produces substantial changes in the dynamical landscape of the climate attractors.
In particular, the extent of the temperate branch is significantly reduced: for the fully land-covered planet, temperate states disappear beyond a critical eccentricity of $e\approx0.11$, compared to $e\approx0.20$ for the nominal land planet.

These results support our interpretation that the breakdown of the MFA at high eccentricity is fundamentally tied to the inability of low-inertia climates to efficiently damp seasonal forcing. More broadly, they suggest that the validity of the MFA depends not only on eccentricity, but also on the thermal inertia of the climate system.

\subsection{\label{app-subsec:sensitivity-C-values}Sensitivity to the adopted heat capacities}

The effective heat capacities adopted in Gen-EBM (Table~\ref{app-tab:heat-capacity}) follow \citet{williams1997habitable}, who in turn based their land heat capacity ($C_l$) on \citet{north1983simple}. \citet{williams1997habitable} also noted that the effective heat capacities of the ocean and sea-ice depend unpredictably on atmospheric and oceanic dynamics (for instance, the wind-mixed ocean layer directly affects the assumed ocean heat capacity). To assess the sensitivity of our results to the absolute magnitude of the thermal inertia, we repeat the MFA experiments of Sect.~\ref{subsec:mean-flux-approximation} after uniformly scaling the surface heat capacities in Table~\ref{app-tab:heat-capacity} by factors of 0.5 and 2 relative to the fiducial values. This preserves the relative ratios between land, ocean, ice, and cold ice, while also allowing us to investigate how equilibration time varies depending on the assumed heat capacities.

\begin{figure*}[ht]
\centering

{\bfseries Low heat capacity set: $0.5C_{\rm fid}$}\\
\includegraphics[width=0.88\hsize]{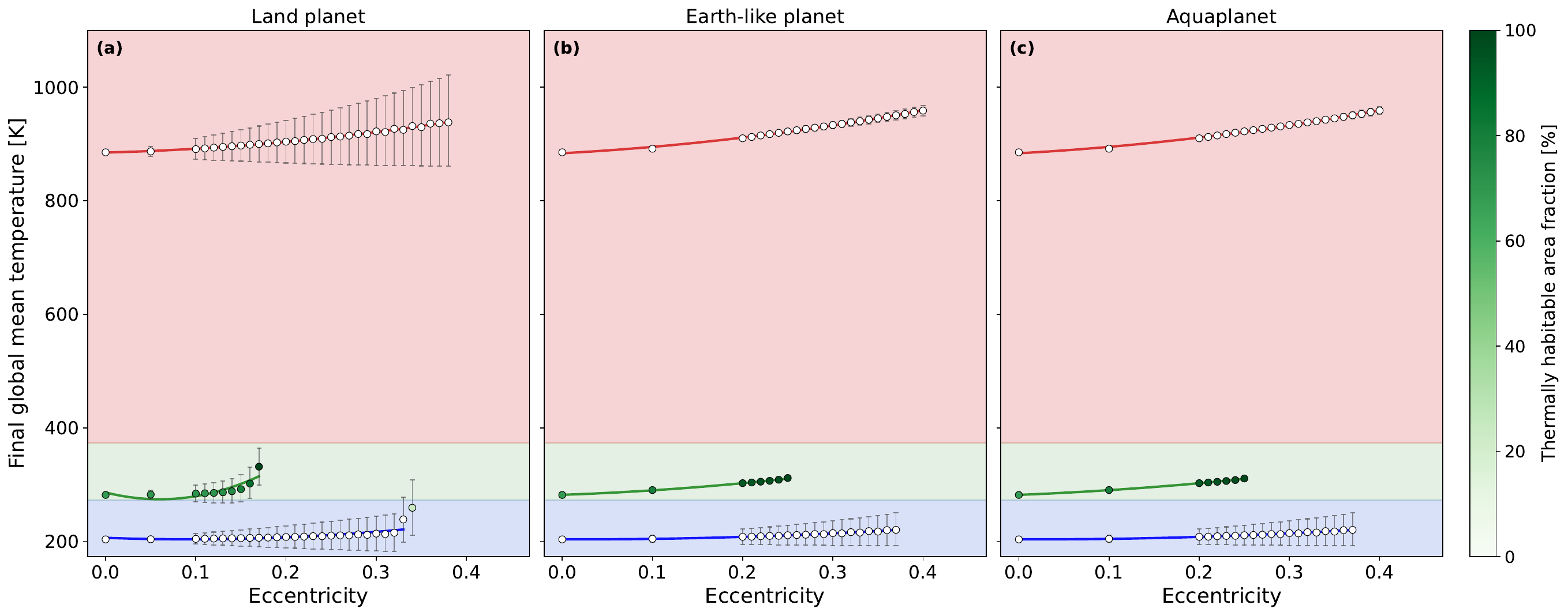}

\vspace{0.1cm}

{\bfseries Fiducial heat capacity set: $C_{\rm fid}$}\\
\includegraphics[width=0.88\hsize]{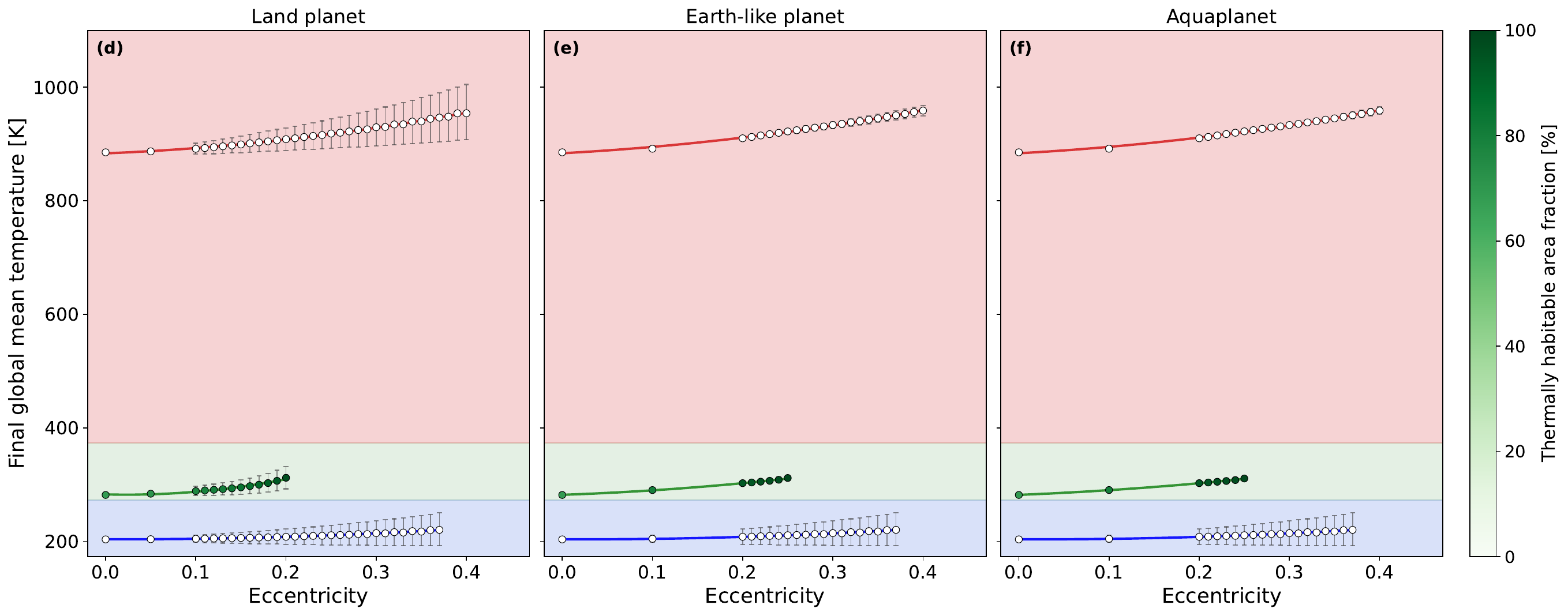}

\vspace{0.2cm}

{\bfseries High heat capacity set: $2C_{\rm fid}$}\\
\includegraphics[width=0.88\hsize]{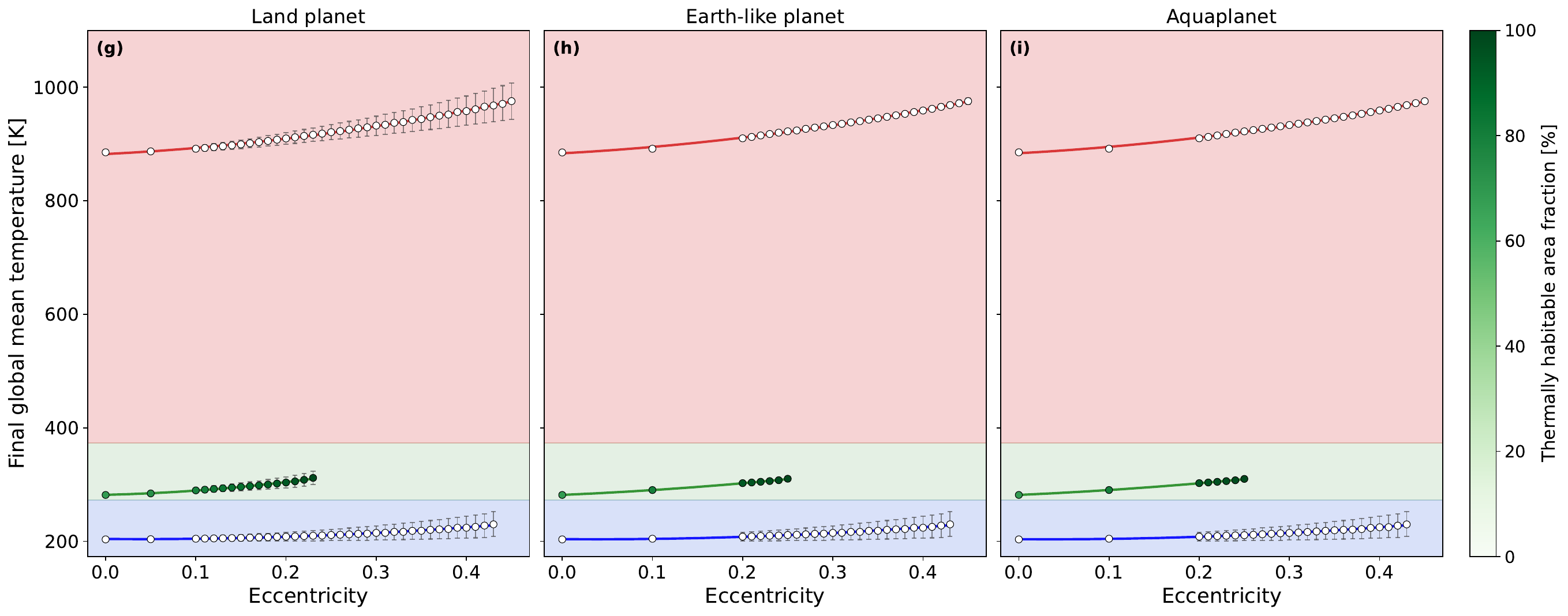}

\caption{
Sensitivity of the MFA bifurcation diagrams to the adopted effective heat capacities. From top to bottom, the surface heat capacities are uniformly scaled by factors of 0.5, 1, and 2 relative to their fiducial values.
}
\label{fig:BD_MFA_heat_capacity_sensitivity}
\end{figure*}

Figure~\ref{fig:BD_MFA_heat_capacity_sensitivity} compares the resulting bifurcation diagrams for the low- ($0.5C_{\rm fid}$), fiducial- ($C_{\rm fid}$), and high-thermal-inertia ($2C_{\rm fid}$) configurations. The strongest sensitivity is seen for the land planet (panel a), for which reducing the heat capacity substantially increases the amplitude of the periastron–apoastron temperature variations. This is particularly apparent for the cold attractor at $e=0.34$, which has a decadal-mean global temperature of approximately 259~K, while its global-mean temperature varies between approximately 211 and 309~K over the final ten orbits. This is a particularly interesting scenario as the climate fluctuates between globally uninhabitable (snowball) and partially habitable states within a single orbit.

Table~\ref{app-tab:heat-capacity-sensitivity} further summarises Fig.~\ref{fig:BD_MFA_heat_capacity_sensitivity} by detailing how the bifurcation thresholds change with the adopted heat capacity. This is particularly apparent for low thermal inertia surfaces (land and ice). For the land planet, the maximum eccentricity supporting a temperate attractor increases from $e=0.17$ at $0.5C_{\rm fid}$ to $e=0.20$ for the fiducial configuration to $e=0.23$ for $2C_{\rm fid}$. In contrast, the corresponding threshold remains at $e=0.25$ for both the Earth-like planet and aquaplanet over the full range of heat capacities tested. The snowball branch is also affected: at high assumed heat capacities, snowball states remain stable to larger eccentricities, reaching $e=0.43$ for all three surface configurations, compared with $e=0.34-0.37$ in the fiducial and low-heat-capacity cases. This extends the range over which snowball and post-runaway attractors coexist, showing that increasing the thermal buffering of a planet can extend the range over which climate multistability can occur.

\begin{table}[ht]
\caption{\label{app-tab:heat-capacity-sensitivity}
Sensitivity of the climate attractor structure and equilibration times to the adopted effective heat capacities.}
\centering
\begin{tabular}{lccc}
\hline\hline
 & \textbf{Land} & \textbf{Earth-like} & \textbf{Aquaplanet} \\
\hline

\multicolumn{4}{c}{$C/C_{\rm fid}=0.5$} \\
\hline
$e_{\rm temperate}^{\max}$ & 0.17 & 0.25 & 0.25 \\
$e_{\rm snowball}^{\max}$ & 0.34 & 0.37 & 0.37 \\
$t_{\rm eq}^{\max}$ [yr] & 20 & 100 & 150 \\

\hline
\multicolumn{4}{c}{$C/C_{\rm fid}=1.0$} \\
\hline
$e_{\rm temperate}^{\max}$ & 0.20 & 0.25 & 0.25 \\
$e_{\rm snowball}^{\max}$ & 0.37 & 0.37 & 0.37 \\
$t_{\rm eq}^{\max}$ [yr] & 40 & 100 & 150 \\

\hline
\multicolumn{4}{c}{$C/C_{\rm fid}=2.0$} \\
\hline
$e_{\rm temperate}^{\max}$ & 0.23 & 0.25 & 0.25 \\
$e_{\rm snowball}^{\max}$ & 0.43 & 0.43 & 0.43 \\
$t_{\rm eq}^{\max}$ [yr] & 60 & 200 & 320 \\

\hline
\end{tabular}

\tablefoot{
$e_{\rm temperate}^{\max}$ and $e_{\rm snowball}^{\max}$ denote the largest eccentricities at which the temperate and snowball attractors, respectively, are found in the simulations. 
$t_{\rm eq}^{\max}$ denotes the maximum equilibration time.
}

\end{table}

Also, as expected from the governing energy-balance equation (Eq.~\ref{eqn:EBM-equation}), increasing the effective heat capacity increases the time required for the climate to adjust to changes in forcing. The maximum equilibration time increases from approximately 20 to 60~years for the land planet, from 100 to 200~years for the Earth-like planet, and from 150 to 320~years for the aquaplanet between the $0.5C_{\rm fid}$ and $2C_{\rm fid}$ experiments.

Overall, these sensitivity experiments show that the exact bifurcation thresholds, orbital temperature amplitudes, and equilibration times depend quantitatively on the adopted effective heat capacities. However, the main qualitative result remains unchanged: land-dominated climates respond more strongly to eccentric forcing, whereas ocean-rich planets are thermally-buffered by their larger thermal inertia.

\end{appendix}
\end{document}